\documentclass[twocolumn, superscriptaddress, pra, amsmath,amssymb, aps, floatfix, 10pt]{revtex4-2}

\usepackage{graphicx}
\usepackage{dcolumn}
\usepackage{bm}
\usepackage{dutchcal}
\usepackage[protrusion=true, expansion=true]{microtype}

\usepackage{braket}
\usepackage{bm,xcolor}
\usepackage{appendix}
\usepackage{amsmath}
\usepackage{amsfonts}
\usepackage{amssymb}
\usepackage{extarrows}
\usepackage{xfrac}
\usepackage{mathrsfs}

\usepackage{siunitx}
\DeclareMathOperator{\tr}{tr}

\definecolor{navyblue}{rgb}{0.0, 0.0, 0.5}
\usepackage[colorlinks=true, breaklinks=false, linkcolor=navyblue, urlcolor=navyblue, citecolor=navyblue]{hyperref}

\begin{document}

\title{Passive leakage removal unit based on a disordered transmon array}

\author{Gonzalo Martín-Vázquez}
\thanks{These two authors contributed equally}
\affiliation{Nano and Molecular Systems Research Unit, University of Oulu, FI-90014 Oulu, Finland}
\affiliation{Departamento de Física Aplicada II, Universidad de Sevilla, E-41012 Sevilla, Spain}
\author{Taneli Tolppanen}
\thanks{These two authors contributed equally}
\affiliation{Nano and Molecular Systems Research Unit, University of Oulu, FI-90014 Oulu, Finland}
\author{Matti Silveri}
\affiliation{Nano and Molecular Systems Research Unit, University of Oulu, FI-90014 Oulu, Finland}

\date{\today}

\begin{abstract}
Leakage out from the qubit subspace compromises standard quantum error correction protocols and poses a challenge for practical quantum computing. We propose here a passive leakage removal unit based on an array of coupled disordered transmons and last-site reset by feedback-measurement or dissipation. We find that the unit  effectively removes leakage with minimal effect to the qubit subspace by taking advantage of energy level disorder for qubit subspace and resonant leakage energy levels. There are two optimal measurement rates for removing leakage, which we show analytically to correspond the characteristic time scales of the propagation and disintegration of the leakage excitation. The performance of the leakage removal unit depends on the strength of disorder, coupling between the transmons, and the length of the array. We simulate the system under experimentally feasible parameters and present an optimal configuration. Our approach is readily compatible with existing superconducting quantum processors considering realistic conditions.
\end{abstract}

\maketitle

\section{Introduction} 
In many quantum computing platforms, a qubit is defined as a two-level subspace of a larger Hilbert space of a physical device~\citep{nielsen10}. These systems have a finite probability to be excited out of the subspace, a phenomenon known as leakage~\citep{fazio1999}. Leakage causes correlated errors that accumulate and propagate among qubits, challenging the effectiveness of quantum error correction for fault-tolerant quantum computing~\cite{ghosh2013}. Even in mature technological platforms, such as superconducting transmon qubits~\citep{Chow2014, koch2007}, many operations can produce leakage, such as single-qubit and entangling gates~\citep{motzoi2009, Barends2014} and measurements~\citep{sank2016}. Creation of leakage can be reduced by various engineered driving pulse approaches~\cite{motzoi2009, gambetta2011,Jun15, hyyppa2024}.

Removing leakage needs inevitably extra hardware or additional pulses or measurements which necessarily creates additional qubit subspace errors. This trade-off is in general tolerable as qubit errors are correctable via standard quantum error correction~\cite{Gottesman09} but leakage errors are not. Leakage removal approaches can be grouped into three different types: i)~\textit{Swap},~in which additional operations are included to regularly swap the role of data and auxiliary qubits~\citep{fowler2013, ghosh2015, brown2019, McEwen2021}; ii)~\textit{Feedback},~in which errors are identified by first measuring and then correcting by applying feedback to the qubit to return it to the computational subspace~\citep{bultink2020, varbanov2020}; and iii)~\textit{Direct},~in which an additional operation or hardware is implemented to transfer the leakage excitation without disturbing the computational states~\citep{battistel2021, marques2023, Miao2023}. Typical leakage removal approaches are dynamic involving either pulses, measurements or on-off couplings, increasing complexity and operation times as they cannot be executed parallel with  algorithms or error-correction cycles.
 
\begin{figure}
\includegraphics[width=\linewidth]{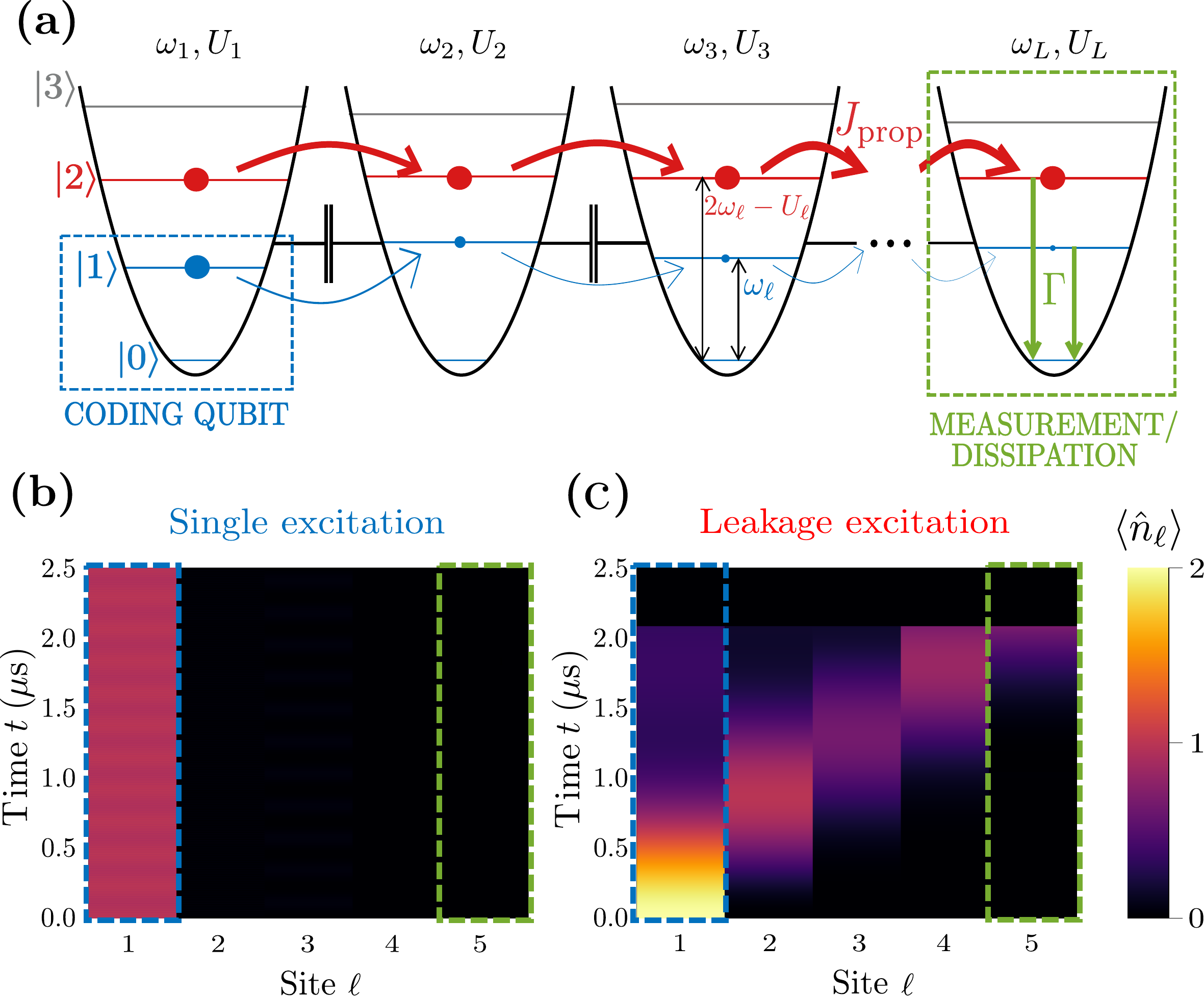}
\caption{\label{fig:LRU_scheme} Schematic of the leakage removal unit protocol. (a)~Transmon array with disordered frequency $\omega_\ell$ and anharmonicity $U_\ell$ tuned on-resonance for the leakage levels $\ket{2}$ and off-resonance for the qubit subspace levels. An example trajectory for transmon population as function of time and site for an initial qubit (b) and leakage (c) excitations. The parameters are: average on-site energy $\bar \omega/2\pi =$~\SI{7.5}{\giga\hertz}, average anharmonicity $\bar U/2\pi =$~\SI{250}{\mega\hertz}, disorder $W/2\pi =$~\SI{100}{\mega\hertz}, and nearest-neighbor hopping rate $J/2\pi =$~\SI{5}{\mega\hertz}.}
\end{figure}

In this article, we propose a passive leakage removal unit based on a disordered transmon array. It is generalizable to other quantum devices with similar energy level and coupling structures. Its operational principles are: efficient leakage mobility by resonant leakage energy levels, qubit subspace protection via disorder-induced localization, and optimized leakage removal by feedback measurement or passive dissipation far from the coding qubit. An essential feature is that both transmon frequency and anharmonicity have site-to-site disorder. We show that intentional anharmonicity disorder in the range of \SIrange{10}{100}{\mega\hertz} or larger opens new useful engineering capabilities. Transmon anharmonicity originates from the internal capacitance, and its value can be selected in fabrication with good accuracy. In-situ tunable frequency disorder is a widely used method for addressing and controlling individual transmon sites~\cite{Krantz19, Saxberg22}. Here frequency and anharmonicity disorder profiles are matched so that leakage energy levels are in resonance but qubit subspaces are off resonance. Then a leakage excitation moves resonantly while single excitations remain localized at the coding qubit site. By applying a feedback measurement or passive dissipation on the edge site at a specific rate, we can selectively remove leakage from the system. Our proposal is akin to the direct type leakage removal units~\cite{battistel2021, marques2023, Miao2023} with the key distinction lying in its passive nature or to high-order filters~\cite{thorbeck24} with the difference that our proposal utilizes just regular transmon qubits without extra specialized hardware.

The structure of the article is as follows. In Sec.~\ref{model}, we introduce the physical model for the LRU. Section~\ref{numerical_results} and Appendix~\ref{app:alternative_reset_methods} evaluate the LRU performance across different parameters and conditions based on numerical simulations. In Sec.~\ref{analytical_results} and App.~\ref{app:analytical_model}, we offer a physical interpretation through the study of an analytical model. Comprehensive mathematical derivations are given for the feedback measurement and the dissipation reset case in Apps.~\ref{app:fb_analytics}-\ref{app:dissipation_analytics}. Appendix~\ref{app:minimal_leakage_removal} summarizes the main analytical results for the simplest scenario. Finally, Section~\ref{discussion} presents a discussion on the feasibility of the results and compares them with other LRU methods, while Section~\ref{conclusion} provides the concluding remarks.

\section{disordered transmon array with last-site reset by feedback measurements or dissipation}\label{model}
The leakage removal unit (LRU) we propose is based on the unitary time evolution of excitations in an array of transmons with inhomogeneous parameters, and a non-unitary interaction via feedback measurements or engineered dissipation that removes excitations far away from the coding qubit, see Fig.~\ref{fig:LRU_scheme}(a). The unitary time evolution of an array of $L$ transmons is given by the Bose-Hubbard Hamiltonian~\citep{hacohen-gourgy15, roushan17}
\begin{equation}
\frac{\hat{H}_{\rm{BH}}}{\hbar} = \sum_{\ell=1}^L \left[\omega_{\ell} \hat{n}_\ell- \frac{U_{\ell}}{2} \hat{n}_{\ell} (\hat{n}_{\ell}-1)+J_{\ell} \left(\hat{a}_{\ell}^\dagger \hat{a}^{}_{\ell+1}+\textrm{h.c.} \right)\right],
\label{Hamiltonian_BH}
\end{equation}
where $\hat{a}_\ell$ represents the bosonic annihilation operator at site $\ell$, while $\hat{n}_\ell$ denotes the corresponding number operator. The nearest-neighbor hopping rate $J_\ell \equiv J$ is assumed constant but the values of the on-site energy $\omega_\ell$ and the anharmonicity $U_\ell$ vary between transmons. The anharmonicity $U_\ell$ is assumed to have large disorder profile $U_\ell \sim \bar U+[-W,W]$ with disorder strength $W$ realized in fabrication and the values of $\omega_\ell$ can be locally adjusted by applying magnetic flux through the transmons. 

For the LRU, we choose $\omega_\ell$ such that the energy of the second excited state, the leakage excitation $\ket{2}$, is uniform across the array, $E_2 \equiv E_2^\ell  =\hbar(2\omega_\ell- U_\ell)$ for all $\ell$, and the energy of the first excited level is inhomogeneous $E_{1}^{\ell}=\hbar \omega_\ell$. Since the inhomogeneities in both parameters are interconnected, we will generally refer to them as \textit{disorder}. In this manner, a leakage excitation at the first site (the coding qubit site) will resonantly propagate to the last site (the measured/dissipated site), while a single excitation will remain localized at the qubit site, see Fig.~\ref{fig:LRU_scheme}(b-c). For a ratio~$J/\bar{U} \ll 1 $, this behavior can be interpreted using an effective model where leakage excitations move with a hopping rate~$J_{\rm prop}=2J^2/\bar{U}$~\citep{mansikkamaki21, mansikkamaki22}. Accordingly, we can expect that the time it takes for a leakage excitation to move between adjacent sites is $T_{\rm prop}=\pi/(2J_{\rm prop})$, see Fig.~\ref{fig:LRU_scheme}(b). At the last site, we implement feedback measurement at the rate $\Gamma_{\rm fb}$ which resets the transmon to the ground state. The feedback measurement process involves measuring the system, evaluating the classical outcome, and applying a conditional gate. Alternatively, we can introduce a dissipative process removing excitations at the rate $\Gamma_{\rm d}$.

\begin{figure}
\includegraphics[width=1.0\linewidth]{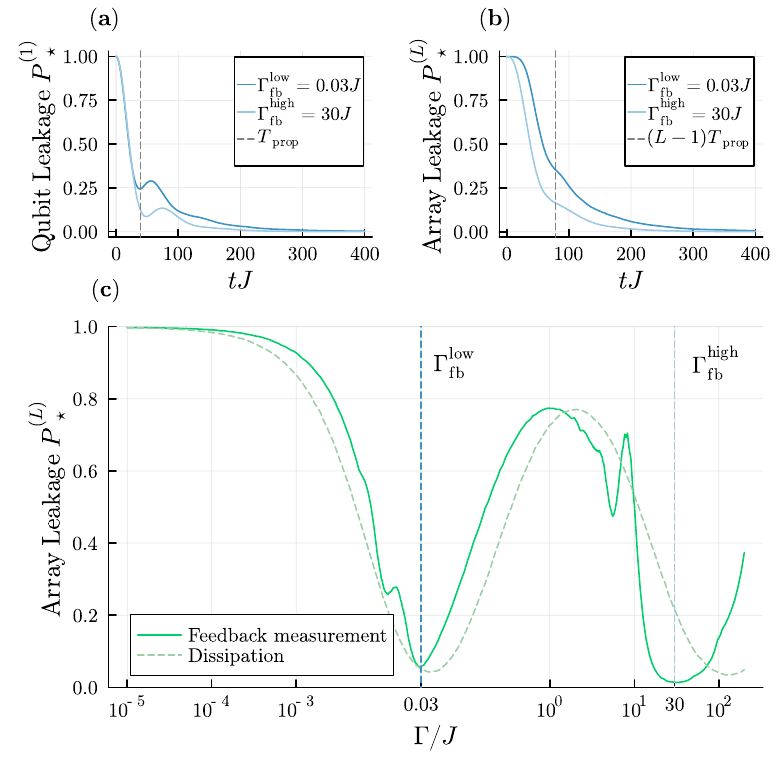}
\caption{\label{fig:leakage_works} Dependence of the leakage population on the measurement and dissipation rates. Time evolution of the leakage population at the qubit site~(a) and in the whole array~(b) for low~(light blue) and high~(dark blue) measurement rates. (c)~Leakage population in the whole array at the final time $tJ=200$ as a function of the feedback measurement rates (green solid line) and dissipation rate (green dashed line). Vertical blue dashed lines indicate the two optimal measurement rates $\Gamma_{\rm fb}^{\rm{low}}=0.03J$ and $\Gamma_{\rm fb}^{\rm{high}}=30J$ for the feedback measurement. For dissipation we find two optimal rates $\Gamma_{\rm d}^{\rm{low}}=0.04J$ and $\Gamma_{\rm d}^{\rm{high}}=130J$. The parameters are same as in Fig.~\ref{fig:LRU_scheme}~with~$L=3$.}
\end{figure}

\begin{figure*}
\includegraphics[width=1.0\linewidth]{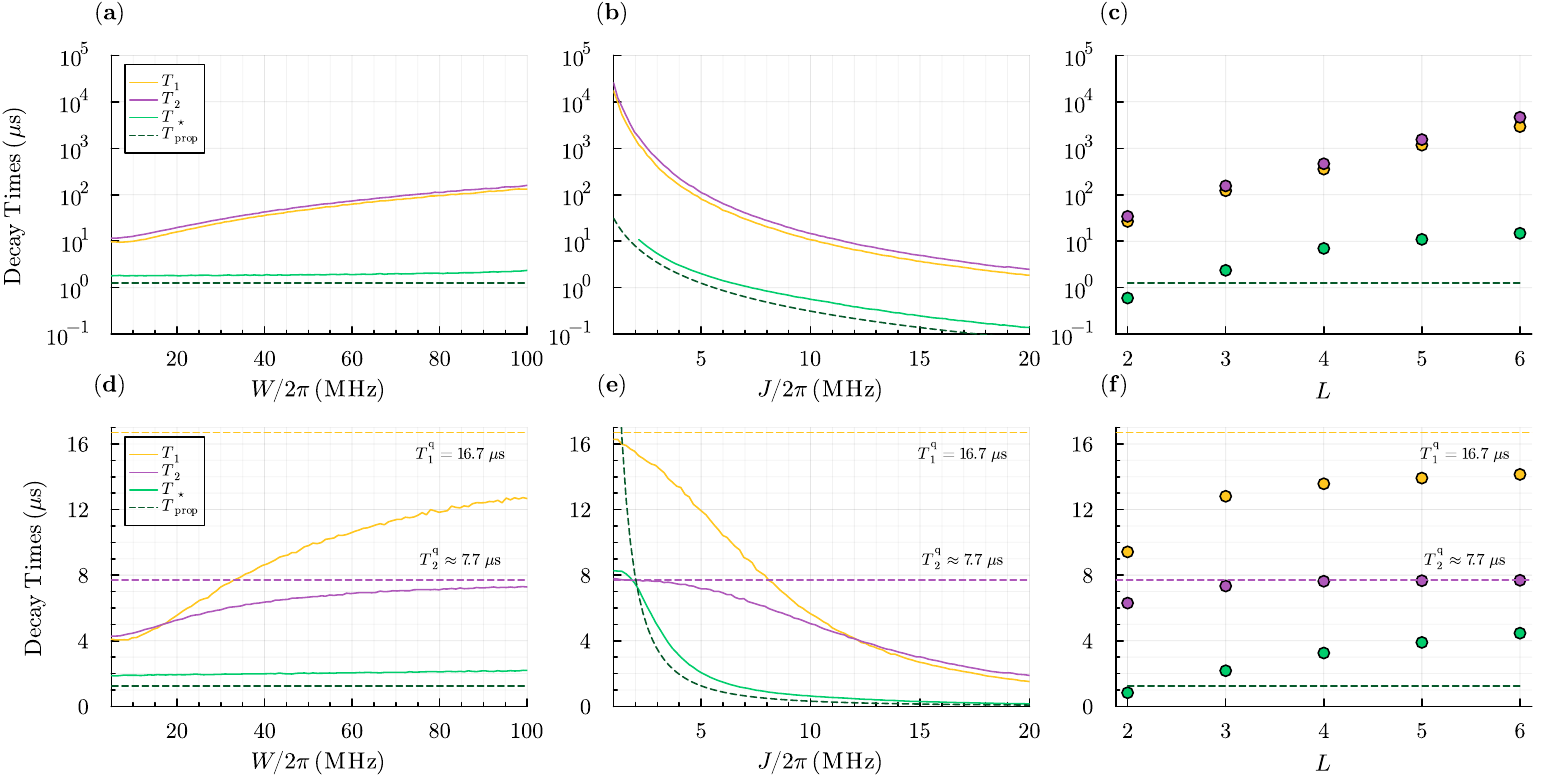}
\caption{\label{fig:leakage_qubit_effect_realistic}~LRU performance for the low optimal feedback measurement rate~$\Gamma_{\rm fb}^{\rm{low}}$: leakage population decay time $T_{\star}$, leakage population propagation time $T_{\rm prop}$, qubit dissipation time~$T_1$, and qubit decoherence time~$T_2$ under ideal (upper panels) and experimentally relevant (bottom panels) conditions. Decay and propagation times as a function of: (a,d) disorder $W$ for $J/2\pi=$~\SI{5}{\mega\hertz} and $L=3$, (b,e) nearest-neighbor hopping rate~$J$ for $W/2\pi=$~\SI{100}{\mega\hertz} and $L=3$, and (c,f) array length $L$ for $W/2\pi=$~\SI{100}{\mega\hertz} and $J/2\pi=$~\SI{5}{\mega\hertz}. For the experimental conditions, we used relaxation time $T^{\rm q}_1 =$~\SI{16.7}{\micro\second}, pure dephasing time ${T}_\phi^\text{q} = $~\SI{10}{\micro\second} yielding decoherence time ${T}^{\rm q}_2 \approx$~\SI{7.7}{\micro\second}, and temperature $T=$~\SI{100}{\milli\kelvin}. In panel (b) the leakage population decay times $T_\star$ are omitted for $J/2\pi\leq$~\SI{2}{\mega\hertz} due to the fitting not being reliable.}
\end{figure*}

In numerical simulations, we model transmons as qutrits allowing to evaluate the performance of leakage removal through the leakage population $P_\star^{(L)}(t) = \braket{\sum_{\ell=1}^L \hat{n}_\ell(\hat{n}_\ell-\hat{I})/2}$. 
First, from the point of view of the coding qubit, we see in Fig.~\ref{fig:leakage_works}(a) that the local leakage population $P_\star^{(1)}(t)$ is initially removed in the time scale~$T_{\rm prop}$ independent on the measurement rate. When the measurement rate is chosen optimally, the late time return revivals are effectively damped. Second, Fig.~\ref{fig:leakage_works}(b) shows the dynamics of the leakage population of the entire array $P_\star^{(L)}(t)$, demonstrating transport through the array and then exponential decay of the leakage population by measurements/dissipation at the last site. By mapping a wide range of measurement rates, we identify two optimal rates $\Gamma_{\rm fb}^{\rm{low}}$ and $\Gamma_{\rm fb}^{\rm{high}}$ where the removal of the leakage population is the most effective and fastest, see green solid line Fig.~\ref{fig:leakage_works}(c). Similarly for dissipation at the last site, two optimal rates are observed, see green dashed line in Fig.~\ref{fig:leakage_works}(c). Importantly, implementing dissipation makes the LRU fully passive and less restrictive in terms of rate requirements, though it necessitates engineering dissipation at the edge site implemented for example by coupling to a lossy resonator~\cite{Ma19} or an engineered quantum circuit refrigerator~\cite{Silveri17, Silveri19}.

The feedback measurement process takes currently approximately \SIrange{1.3}{1.6}{\micro\second}~\cite{baumer2024, Baumer24PRL, Song24}. Thus, in the next section, for the feedback measurement, we will focus on the low optimal feedback measurement rate $\Gamma_{\rm fb}^{\rm low}/J = 0.03$, as this rate is within the range of current and near-future experimental feasibility with $J/2\pi\sim\SIrange{1}{10}{\mega\hertz}$. In contrast, the dissipation frequency could, in principle, be designed to match both optimal rates, so for the dissipation reset we will study both the low $\Gamma_{\rm d}^{\rm{low}}/J=0.04$ and high $\Gamma_{\rm d}^{\rm{high}}/J=100-130$ optimal rates.

\section{Performance of the leakage removal unit}\label{numerical_results}
\begin{figure*}
\includegraphics[width=1.0\linewidth]{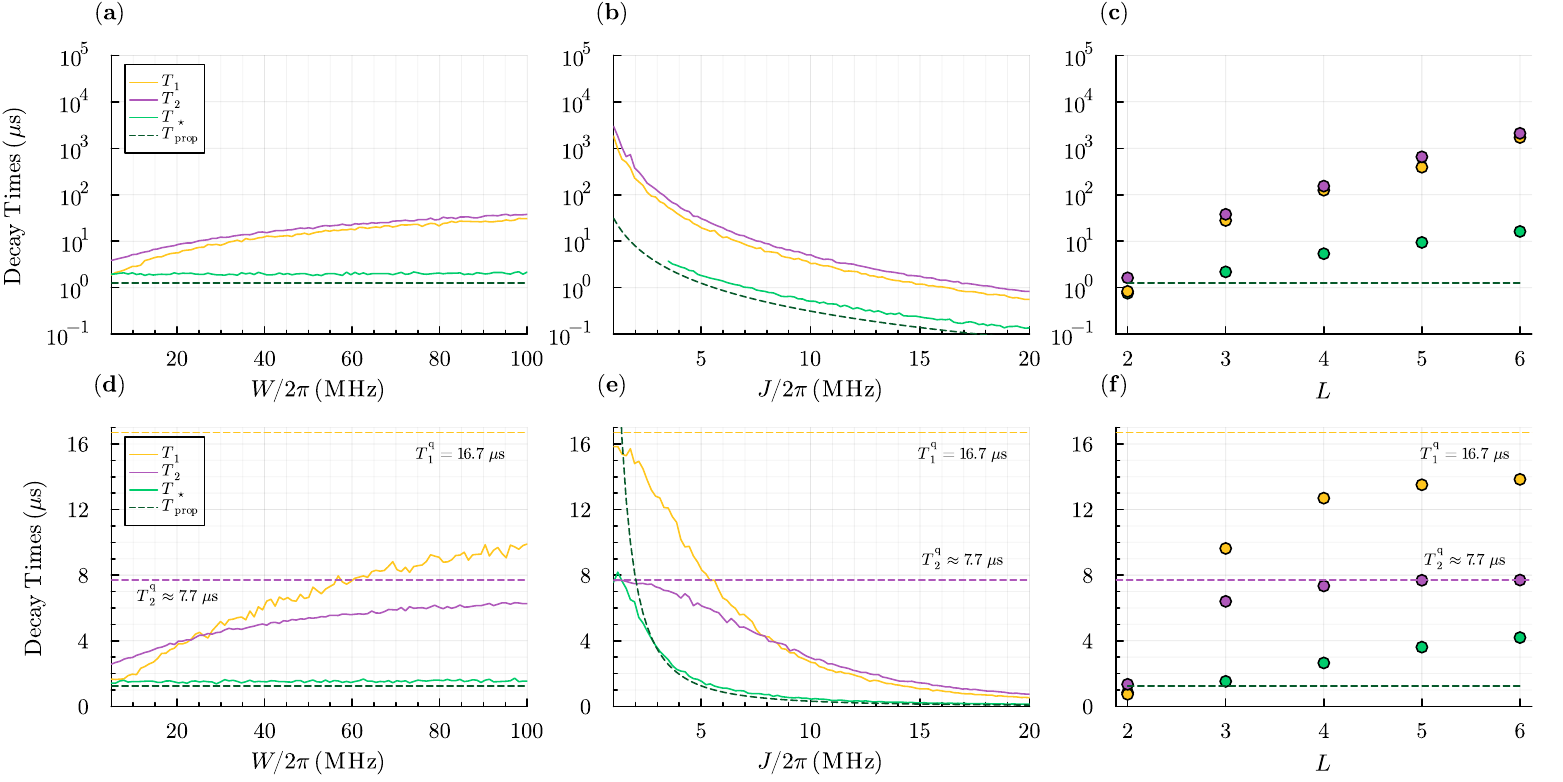}
\caption{\label{fig:decay_diss_high}
   ~LRU performance for the high optimal dissipation rate~$\Gamma_{\rm{d}}^{\rm{high}}$: leakage population decay time $T_{\star}$, leakage population propagation time $T_{\rm prop}$, qubit dissipation time~$T_1$, and qubit decoherence time~$T_2$ under ideal [panels (a)-(c)] and experimentally relevant conditions [panels (d)-(f)]. We use the same parameters as in Fig.~\ref{fig:leakage_qubit_effect_realistic} and $\Gamma_{\rm{d}}^{\rm{high}} = 2 \overline{U} = 100J = 2\pi\times$~\SI{500}{\mega\hertz}.
}
\end{figure*}

We quantify the performance of the LRU determining the decay time $T_{\star}$ for the leakage population in the whole array, as well as, the dissipation time $T_1$ and the decoherence time $T_2$ at the first site.  The times are determined from fits to exponential decay to the late time dynamics after the initial dynamics beyond $T_{\rm prop}$, see~Fig.~\ref{fig:fig3_fits} of App.~\ref{app:numerical_simulation}. We study these decay times as a function of the disorder strength~$W$, the nearest-neighbor hopping rate~$J$, and the array length~$L$, see Figs.~\ref{fig:leakage_qubit_effect_realistic}-\ref{fig:decay_diss_high}.

\subsection{Last-site reset by periodic feedback measurement}
Qubit dissipation and decoherence times $T_1$ and $T_2$ increase polynomially with increasing disorder strength~$W$, while the leakage population removal time $T_{\star}$ remains essentially unaffected, see~Fig.~\ref{fig:leakage_qubit_effect_realistic}(a). This is because the disorder only increases the energy differences of the first excited states, $E_1^\ell$, without influencing the second energy level difference, $E_2$. The removal of the leakage population from the first site depends only on its mobility. Specifically, it takes approximately a time $T_{\rm prop}=\pi/2J_{\rm prop}$ for the leakage population to move from the first site to the second site, excluding boundary effects~\cite{mansikkamaki22}, see Fig.~\ref{fig:leakage_works}(a). This also accounts for the initial delay by $(L-1)T_{\rm prop}$ observed in the leakage population of the whole array, see~Fig.~\ref{fig:leakage_works}(b). Hopping time $T_{\rm prop}$ can be reduced by increasing the ratio~$J/\bar{U}$, although it also reduces the hopping times of single excitations in the same way, see~Fig.~\ref{fig:leakage_qubit_effect_realistic}(b).

As a function of the array length $L$, dissipation and dephasing times $T_1$ and $T_2$ increase exponentially while $T_{\star}$ increases linearly, see Fig.~\ref{fig:leakage_qubit_effect_realistic}(c). In a longer array, due to scattering and imperfect transport, the probability distribution of a leakage excitation spreads across the array reducing the efficiency of removal at the last site. Under parametric disorder, the qubit eigenstate $\ket{\widetilde 1}_1$ on the first site is localized with exponentially decaying contributions from other sites $\ell$ proportional to $\exp(-\ell/\xi)$ where $\xi$ is the localization length~\cite{Anderson58}. As the array is open only at the edge $L-1$ sites displaced from the coding qubit, the effective qubit dissipation and decoherence scale as $\exp(L/\xi)$. In summary, effective qubit protection is achieved by short localization length~$\xi$ via large disorder strength or by having a long array. 

Next, we perform a similar analysis to evaluate the LRU performance under experimentally feasible, state-of-the-art conditions. We take into account that transmons operate at non-zero temperatures and undergo dephasing and dissipation processes, see Figs.~\ref{fig:leakage_qubit_effect_realistic}(d)-(f). We consider the following parameters: relaxation time $ T_1^{\rm q}=$\SI{16.7}{\micro\second}, pure dephasing time $T_{\phi}^{\rm q} =$\SI{10}{\micro\second} yielding a decoherence time $T_{2}^{\rm q} \approx$\SI{7.7}{\micro\second}~\citep{Rosen2024}, and temperature $T=$\SI{100}{\milli\kelvin}~\citep{jin2015}. This establishes an upper limit for the LRU performance, given by $T_1 \leq T_1^{\rm q}$ and $ T_2 \leq T_2^{\rm q}$, see dashed lines in Fig.~\ref{fig:leakage_qubit_effect_realistic}(d)-(f). In this scenario, increasing the disorder strength causes $T_1$ and $T_2$ to rise polynomially while keeping $T_{\star}$ nearly constant, similar to the ideal case, see Fig.~\ref{fig:leakage_qubit_effect_realistic}(d). Increasing $J$ reduces decay times, causing faster leakage removal but also more qubit degradation.  We find $J/2\pi =$~\SI{5}{\mega\hertz} optimal here in the sense that leakage reduction occurs fast while reduction in coherence times is relatively small. Finally, all decay times increase as a function of the array length~$L$, similar to the ideal case, see Fig.~\ref{fig:leakage_qubit_effect_realistic}(f). Notably, $T_2^{\rm q}$ is less affected than $T_1^{\rm q}$ in all cases. Based on these findings, we can estimate an optimal set of parameters to be $W/2\pi=$\SI{100}{\mega\hertz}, $J/2\pi=$\SI{5}{\mega\hertz}, and $L=3$, which results in $T_1\approx$~\SI{12.67}{\micro\second}, $T_2\approx$~\SI{7.29}{\micro\second} ($T_1 \approx 0.76 \, T_{1}^{\rm q}$, and $T_2 \approx 0.95 \, T_{2}^{\rm q}$), and $T_{\rm prop}=$~\SI{1.25}{\micro\second} and $T_{\star}\approx$~\SI{2.2}{\micro\second}.

Charge noise in transmon devices causes dispersions in the higher excited states. However, with current state-of-the-art transmon devices $E_{\rm J}/E_{\rm C}$ fraction has values approx $E_{\rm J}/E_{\rm C}=70-75$ with $E_{\rm C}/2\pi$ having values from~\SIrange{200}{300}{\mega\hertz}~\cite{Rosen2024, karamlou24, acharya23} resulting the charge dispersion of the second excited level $\epsilon_2/2\pi \approx$~\SI{20}{\kilo\hertz}, which is small compared to the effective coupling between the energy levels $J_{\rm prop}/2\pi\approx$~\SI{200}{\kilo\hertz}. Thus, the charge noise causes no impediment for the leakage propagation.

\subsection{Last-site reset by engineered dissipation}
For the low optimal dissipation rate we find similar results as for the low optimal feedback measurement rate, see Fig.~\ref{fig:decay_diss_low} and App.~\ref{app:numerical_simulation}. We expect that the optimal parameters found for the feedback measurement apply for dissipation with similar performance.

For the high optimal dissipation rate, see Fig.~\ref{fig:decay_diss_high}, in general we find an improvement in leakage removal times at the cost of shorter dissipation $T_1$ and dephasing times~$T_2$ compared to the case of the low optimal feedback measurement rate. Comparing Fig.~\ref{fig:leakage_qubit_effect_realistic}(d) and Fig.~\ref{fig:decay_diss_high}(d) at disorder $W/2\pi =$~\SI{100}{\mega\hertz}, the leakage removal time $T_\star$ is reduced by roughly \SI{1}{\mu\second}, representing a $100\%$ improvement, while the dissipation time $T_1$ and dephasing time $T_2$ are reduced by \SI{3}{\mu\second} and \SI{1}{\mu\second}, corresponding to a $23\%$ and $13\%$ worsening, respectively. Since the leakage removal time is already about an order of magnitude smaller than the dissipation and dephasing times for the low optimal feedback measurement rate, we do not expect the slightly improved leakage removal time to justify the accompanying degradation in dissipation and dephasing times. We can use a longer array to have comparable qubit protection as with feedback measurement, see Fig.~\ref{fig:leakage_qubit_effect_realistic}(c, f) and Fig.~\ref{fig:decay_diss_high}(c, f), but at these lengths the improvement in leakage removal time is lost. Furthermore, for array length $L = 2$ the high optimal dissipation rate removes qubit population as fast as the leakage population, in contrast to the low optimal rates.

\section{Interplay of leakage excitation propagation and disintegration with measurements and dissipation}\label{analytical_results}
The LRU relies on the resonant propagation of the leakage excitation while localizing qubit subspace excitations at the coding qubit site, as well as the existence of two optimal rates for both feedback measurement and engineered dissipation at the last site. Here we analytically study these phenomena considering the minimal LRU consisting of two transmons in ideal conditions, with excitations located at the first site and the measurement or dissipation at the second site. For a comprehensive explanation of the analytical model and results, see Apps.~\ref{app:analytical_model}-\ref{app:dissipation_analytics}. 

\subsection{Random feedback measurements}
We address the effect of the measurement by considering the master equation where measurements occur randomly at a rate $\Gamma_{\rm fb}$~\citep{cresser06, fuji20, yamamoto2023, zhou2023},
\begin{equation}
    \frac{d\hat{\rho}}{dt}=-\frac{i}{\hbar}[\hat{H}_{\rm{BH}}, \hat{\rho}  ] +\Gamma_{\rm fb} \left( \sum_{n=0}^{2} \hat{\Pi}_{n} \hat{\rho} \hat{\Pi}_{n}^\dagger-\hat{\rho}\right),
    \label{master_equation_LRU_measurement}
\end{equation}
where $\hat{\Pi}_{n}=\hat{I} \otimes \ket{0}\bra{n}$ are the projectors describing the feedback measurement on the second site~\citep{martinvazquez23}. Since the probability of performing a measurement is independent of the state of the system, given by $p_{\textnormal{fb}}=1-e^{-\Gamma_{\rm fb} t}$ over a time interval of length $t$, and the Lindblad operators consist of measurement projectors, the probability of applying different operators $\hat{\Pi}_{n}$ depends solely on the Born's rule probability of the system.

First, we focus on the dynamics and removal of the leakage population. To study the low optimal rate $\Gamma_{\rm fb}^{\rm{low}}$, we consider an effective model given by the perturbative Hamiltonian 
\begin{equation}
\frac{\hat{H}_{\rm{BH}}^{\rm{prop}}}{\hbar} = J_{\rm prop} \left[\hat{n}^\alpha_1+\hat{n}^\alpha_L-\sum_{\ell=1}^L (\hat{\alpha}_{\ell}^\dagger \hat{\alpha}^{}_{\ell+1}+\textrm{h.c.} ) \right],
\label{Hamiltonian_BH_prop}
\end{equation}
which describes the dynamics of leakage excitations propagating as a single particle when $J/\bar{U} \ll 1$~\citep{mansikkamaki22}. Here $\hat{\alpha}_\ell$ represents leakage excitation annihilation operator at site~$\ell$ and $\hat{n}_\ell^\alpha$ is the corresponding number operator.  By solving Eq.~\eqref{master_equation_LRU_measurement} perturbatively, we find the leakage population to decay as
\begin{align}
    P_{\star,\rm low}^{(L)}(t) \approx 
    \begin{cases}
        \rm{e}^{-\frac{\Gamma_{\rm fb}}{2}t },& \Gamma_{\rm fb} \ll 2J_{\rm prop} \\
        \rm{e}^{-\frac{2J_{\rm prop}^2}{\Gamma_{\rm fb}}t},& \Gamma_{\rm fb} \gg 2J_{\rm prop}
    \end{cases}.
    \label{eq:fb_low_cases}
\end{align}
By combining these limits, we obtain the approximate function
\begin{equation}
    P_{\star,\rm low}^{(L)}(t) \approx \exp\left(-\frac{2J_{\rm prop}^2 \Gamma_{\rm fb} }{4J_{\rm prop}^2+\Gamma_{\rm fb}^2}t\right).
    \label{fb_low_combination}
\end{equation}
This result yields the low optimal measurement rate $\Gamma_{\rm fb}^{\rm{low}} \approx 2J_{\rm prop}$ with a decay time of $T_{\star}^{\rm{low,\ fb}} \approx 2/J_{\rm prop}$ coinciding with the effective hopping rate of the leakage excitation $\omega_{\rm prop}=2J_{\rm prop}$.

To study the high optimal rate $\Gamma_{\rm fb}^{\rm{high}}$, we map the dynamics of two transmons to an effective two-level system, whose two states are $\ket{11}$ and the subspace spanned by $\ket{20},\ket{02}$. We analyze the Liouvillian $\mathcal{L}$ of Eq.~\eqref{master_equation_LRU_measurement} over long but finite periods of time~\cite{macieszczak2016, mingatti2018}. By solving $\partial_t \hat{\rho}(t)=\mathcal{L}\hat{\rho}(t) $ perturbatively, we found the leakage population to decay as 
\begin{equation}
    P_{\star,\textrm{high}}^{(L)}(t) \approx \exp \left(-\frac{4J^2\Gamma_{\rm fb}}{\Gamma_{\rm fb}^2+\bar{U}^2}t\right),
\end{equation}
yielding an optimal measurement rate $\Gamma_{\rm fb}^{\rm{high}} \approx \bar{U} $ with a decay time of $T_\star^{\rm{high,\ fb}}\approx \bar{U}/2J^2$. This value is similar to the frequency of leakage excitation disintegration $\omega_{\rm{dis}} =\sqrt{\bar{U}^2+16J^2} $ when $J/\bar{U} \ll 1$.

We can interpret the two optimal rates as measurements that interact with the two time scales of leakage excitation dynamics: leakage excitation propagating as a whole $\ket{20}\xleftrightarrow{\omega_{\rm{prop}}}\ket{02}$ or disintegrating into individual excitations $\ket{\sfrac{20}{02}}\xleftrightarrow{\omega_{\rm{dis}}}\ket{11}$, where $\omega_{\rm{dis}} > \omega_{\rm{prop}}$. When $ \Gamma_{\rm fb} \sim  \omega_{\rm{prop}}$, the measurements effectively observe the leakage excitation and remove it, and when $ \Gamma_{\rm fb} \sim \omega_{\rm{dis}}$, the measurements observe the leakage excitation disintegrated and remove one of the single excitations $\ket{1}$. Interestingly, even for $J/\bar{U} \ll 1$, where the picture of propagating leakage excitations is effective and the probability of the system being in the intermediate disintegrated state $\ket{11}$ is negligible, there is always a measurement rate $\Gamma_{\rm fb}^{\rm{high}}$ at which the system can be observed in this intermediate state, allowing removal of a leakage excitation.

To evaluate how the LRU affects the qubit subspace we consider the decaying of single excitations. A localized excitation decays as $\braket{\hat{n}_1}\approx e^{-t/T_1^{\rm{fb}}}$, and a superposition state decays as $\braket{\hat{\rho}_+}\approx ( 1+e^{-t/T_2^{\rm{fb}}})/2$, where
\begin{equation}
    T_1^{\rm{fb}} \approx \frac{ \Gamma_{\rm fb}^2+(\omega_1 -\omega_2)^2}{2J^2\Gamma_{\rm fb}},
\end{equation}
and $T_2^{\rm{fb}}=2T_1^{\rm{fb}}$, respectively. Note that in this case, we aim to maximize the values of $T_1^{\rm{fb}}$ and $T_2^{\rm{fb}}$ to increase the survival of the qubit subspace population. This occurs for $|\omega_1 -\omega_2|/J \gg 1$ and $\Gamma_{\rm fb} \ll |\omega_1 -\omega_2|$ or $\Gamma_{\rm fb} \gg |\omega_1 -\omega_2|$. Therefore, the best strategies to avoid affecting the qubit subspace population are to set a large disorder, and choose either $\Gamma_{\rm fb}^{\rm{low}} \ll |\omega_1 -\omega_2| $ or $|\omega_1 -\omega_2| \ll \Gamma_{\rm fb}^{\rm{high}}$.

\subsection{Engineered dissipation} 
In the case of using dissipation as the removal element in the LRU, we observe a similar overall performance as with the feedback measurements, see Fig.~\ref{fig:scheme_result}. The master equation is then given by
\begin{equation}
    \frac{d\hat{\rho}}{d t}=-\frac{i}{\hbar}[\hat{H}_{\rm{BH}}, \hat{\rho}] +\Gamma_{\rm d} \left( \hat{a}_{2} \hat{\rho}\hat{a}_{2}^\dagger-\frac{1}{2}\{\hat{a}_{2}^\dagger \hat{a}_{2}, \hat{\rho} \}\right),
    \label{master_equation_LRU_dissipation}
\end{equation}
which can be solved by averaging the quantum trajectories evolving under $\hat{H}_{\rm{eff}}=\hat{H}_{\rm{BH}}-i\hbar\Gamma_{\rm d}\hat{n}_2/2 $. The reduction of the norm of these quantum trajectories $\mathscr{N}_\psi(t) =\braket{\psi_{\rm{eff}}(t)|\psi_{\rm{eff}}(t)}$ corresponds to an increase in the probability of quantum jumps, i.e., dissipation events, $p_{\rm{d}}(t)=1-\mathscr{N}_\psi(t)$. By studying how the norm decays over time, we can infer the removal of excitations due to dissipation.

To study the low optimal rate, $\Gamma_{\rm d}^{\rm{low}}$, we consider the effective model described by Eqs.~\eqref{Hamiltonian_BH_prop} and \eqref{master_equation_LRU_dissipation}, which in this case can be exactly diagonalized. The exact expression for $\mathscr{N}_\psi(t)$ is a complicated, time-dependent function, see Eqs. \eqref{L2_diss_prop_exact1}-\eqref{L2_diss_prop_exact2}; however, we can proceed as we did for the feedback measurement case in Eqs.~\eqref{eq:fb_low_cases}-\eqref{fb_low_combination} and study both limits $\Gamma_{\rm d} \ll 2J_{\rm prop}$ and $\Gamma_{\rm d} \gg 2 J_{\rm prop}$ perturbatively, resulting in 
\begin{equation}
    \mathscr{N}_{\psi,\rm low}(t) \approx \exp\left(-\frac{2J_{\rm prop}^2 \Gamma_{\rm d} }{2J_{\rm prop}^2+\Gamma_{\rm d}^2}t\right).
\end{equation}
This result yields the low optimal dissipation rate~$\Gamma_{\rm d}^{\rm{low}} \approx \sqrt{2}J_{\rm prop}$ with a decay time of $T_{\star}^{\rm{low,\ d}} \approx \sqrt{2}/J_{\rm prop}$. We further show that, at the low optimal dissipation rate, increasing the chain length leads to longer leakage population removal times and a reduction in the low optimal dissipation rate itself, see Eq. \eqref{L_diss_prop_noedgelocalization} for the dependence on the array length~$L$. Moreover, for chains with $L \geq 3$, the presence of borders produces a similar tendency due to the edge-localization effect, see Eq. \eqref{L3_diss_prop_edgelocalization}.

For the high optimal rate $\Gamma_{\rm d}^{\rm{high}}$, we obtain
\begin{equation}
    \mathscr{N}_{\psi,\rm high}(t) \approx \exp\left(-\frac{8J^2 \Gamma_{\rm d}}{4\bar{U}^2+ \Gamma_{\rm d}^2}t\right),
\end{equation}
which yields $\Gamma_{\rm d}^{\rm{high}} \approx 2\bar{U}$ with a decay time $T_\star^{\rm{high, d}}\approx \bar{U}/2J^2$. For the high optimal rate, we also expect a similar increase in the leakage population removal time with an increase in chain length, as the difference lies solely in the last transmon, i.e., disintegration instead of propagation.

For single excitations, we cannot obtain directly $T_1^{\rm{d}}$ or $T_2^{\rm{d}}$ by analyzing the norm of the quantum trajectories, but we can determine the probabilities of dissipation. We have that for a localized excitation  $\mathscr{N}_\psi(t) \approx e^{- t/\tau_1} $, and for a localized superposition $\mathscr{N}_\psi(t)\approx  (e^{-t/\tau_2}+1)/2 $, where
\begin{equation}
    \tau_1=\tau_2\approx \frac{4\left(\omega_1 - \omega_L \right)^2+ \Gamma_{\rm d}^2}{4FJ^2 \Gamma_{\rm d}},
\end{equation}
and $F=\prod_{n=2}^{L-1}  J^2/\left( \omega_1- \omega_n \right)^2$. Similar to the feedback measurements, we aim to maximize $\tau_1 $ and $\tau_2 $ to increase the survival of the qubit subspace population. This is achieved by introducing a large disorder among all transmons, ensuring that $F \ll 1$, and by choosing either $\Gamma_{\rm d}^{\rm low} \ll 2|\omega_1 -\omega_L| $ or $  2|\omega_1 -\omega_L| \ll \Gamma_{\rm d}^{\rm high}$.

For both feedback measurement and dissipation, there is the Zeno effect when $\Gamma_{\rm d},\Gamma_{\rm fb} \to \infty $. In the case of feedback measurements, increasing $\Gamma_{\rm fb}$ increases the number of measurements that keep the initial state frozen, see Fig.~\ref{fig:leakage_works}(c) at high $\Gamma$. For the case of high dissipation, when $\Gamma_{\rm d} \gg \Gamma_{\rm d}^{\rm{high}}$, the probability of dissipation events approaches zero, as $p_{\rm{d}}(t) \approx 1-e^{- (8J^2 /\Gamma_{\rm d})t}$, as described in~Ref.~\cite{daley14}, see Eq.~\eqref{norm_split_dissipation}. Another interesting aspect of leakage removal is that once the disorder is adjusted to match $E_2$ at each site, it no longer influences the optimal measurement rates or their associated decay times,~see~Eq.~\eqref{Hamiltonian_BH_N2}.

\section{Discussion}\label{discussion}
The leakage removal here is based on selective mobility of excitations achieved by utilizing engineered disorder in transmon anharmonicity and frequency. In contrary to many other leakage removal strategies, our approach is passive and built hardware-efficiently just using transmons.  It involves no gates nor pulses that would prolong for example quantum error correction cycles. Superconducting quantum processors already employ transmons both as computational qubits and tunable couplers. In this sense, a subset of transmons could be dynamically delegated leakage removal purposes. The passive approach is slower compared to active ones. This is however compensated by the fact that the passive LRU is always on, that is, leakage removal times $T_{\rm prop}$ and repetition time of a quantum error correction cycle are of same order,~\SI{1}{\micro\second}. Given similar energy-level structures and tailorable anharmonicity disorder, LRU should work also with other qubit types such as with C-shunted flux~\cite{you_low-decoherence_2007, yan16} or C-shunted fluxonium~\cite{nguyen_high-coherence_2019, kwon_gate-based_2021} qubits. Transmons have relatively low anharmonicity. Indeed, larger anharmonicity $\bar{U}$ allows having larger disorder amplitude $W$, which improves qubit subspace protection,~cf.~Fig.~\ref{fig:leakage_qubit_effect_realistic}(a). Disorder in anharmonicity can also suppress unwanted chaos dynamics in superconducting quantum processors~\cite{Blain25}.

Increasing the size of the LRU significantly improves $T_1$ and $T_2$, although it worsens the leakage removal time $T_{\star}$. The impact on $T_{\star}$ could be mitigated by diminishing localization and scattering effects, through e.g. engineering a specific spatial profile to the hopping terms~$J_\ell$~\citep{Xiang2024}. If the anharmonicity disorder can be designed large enough, then a minimal LRU consists of only two transmons. We have also demonstrated that the LRU functions effectively under state-of-the-art experimental conditions, achieving a low $T_{\star}$ with minimal impact on the qubit dissipation and decoherence times $T_1$ and $T_2$. At optimal parameters of $J/2\pi=$~\SI{5}{\mega\hertz} and $W/2\pi=$~\SI{100}{\mega\hertz}, the qubit decoherence time $T_2$ remains largely unchanged, while qubit dissipation $T_1$ is reduced by one-third. This reduction in $T_1$ is attributed to the intrinsic decoherence caused by the additional transmons.

Our approach is most similar to the \textit{direct} leakage removal type, although unlike other direct methods that require active pulses on the coding site~\cite{Miao2023, marques2023}, our method avoids active intervention. This also contrasts swap-based methods~\cite{McEwen2021}, and feedback-based methods~\cite{bultink2020, varbanov2020}, which apply pulses to the coding sites for swapping and correction. As expected, our passive LRU is slower than active methods since it relies on the natural dynamics of the system. As timescales, we use the propagation time~$T_{\rm prop}$ from the coding transmon to the first idle transmon, estimated as $T_{\rm prop} \approx$ \SI{1.25}{\micro\second}, and the time $T_\star$ for exponential decay of the leakage population in the whole array, estimated as $T_\star\approx$~\SI{2.2}{\micro\second}. Then the time to achieve $90\%$ suppression of leakage population at the first site can be estimated as $T_{\rm tot}\lesssim T_{\rm prop} - T_\star \ln [P_{\star, \rm f }^{(1)}/P_{\star,\rm i}^{(1)}]\approx$~\SI{3.7}{\micro\second}. Here $P_{\star, \rm f/i }^{(1)}$ are final and initial leakage populations in the exponential decay part. 
In contrast, active approaches are faster: direct methods remove $80 \%$ of leakage in $\sim$ \SI{1}{\micro\second}--within a QEC cycle--\cite{Miao2023} and $ 99 \%$ as fast as $\sim$ \SI{0.22}{\micro\second}~\cite{marques2023}; swap-based techniques achieve full removal in $\sim$ \SI{0.25}{\micro\second}~\cite{McEwen2021}; and feedback-based schemes operate concurrently with QEC in $\sim$ \SI{1}{\micro\second}~\cite{bultink2020, varbanov2020}.

A recently proposed passive method removes leakage using a custom high-order band-pass filter, achieving faster removal times of \SI{0.36}{\micro\second}, but requiring a filter per transmon, which limits scalability~\cite{thorbeck24}. In contrast, our LRU is constructed using additional transmons, which are fundamental building blocks in standard superconducting devices, allowing for a modular and scalable implementation. For instance, a 2D grid of coding transmons can resonantly route leakage to boundary-placed LRUs. Another advantage is that since our LRU does not require active operation, it is compatible with other leakage removal methods. Performance may improve further by fabricating transmons with higher anharmonicity allowing greater disorder and increased leakage mobility through larger hopping rates~$J$, thus enhancing leakage removal while preserving the computational subspace. For instance, from the ideal case shown in Fig.~\ref{fig:leakage_qubit_effect_realistic}, we estimate that increasing the anharmonicity $U/2\pi$ by \SI{100}{\mega\hertz} allows us to increase the mobility to $J/2\pi\sim$ \SI{10}{\mega\hertz} while maintaining the same level of qubit protection via increased disorder strength. As a result, the propagation and decay can be reduced to approx $1/3$ of the current values: $T_{\rm prop} \approx$ \SI{0.44}{\micro\second}, $T_\star\approx$~\SI{0.7}{\micro\second}, and $T_{\rm tot}\approx$~\SI{1.2}{\micro\second}.

\section{Conclusion}\label{conclusion}
In summary, we have developed a leakage removal protocol for transmon-based quantum processors utilizing passive quantum dynamics in a disorder system and minimal interaction via feedback measurement, which can be applied to existing devices. By levering thorough understanding of the many-body dynamics between transmons, we found optimal feedback measurement rates. It has been recently pointed out that leakage mobility is a desirable feature for an LRU~\citep{camps2024}; in our case, we provided a concrete recipe through disorder engineer to achieve this without substantially sacrificing qubit subspace properties. We provided also an alternative, fully passive protocol by using dissipation instead of a feedback measurement. This approach would involve coupling a dissipative element, such a lossy resonator, to the last qubit to create effective reset. The main ideas are general enough to be applied in qudit schemes~\citep{basyildiz2023, vezvaee2024}, where it would be necessary to equalize between transmons the energy levels of the corresponding leakage excitation, although a reduction in speed mobility is expected~\citep{mansikkamaki22}. In addition, the passive LRU could be used as a hardware-level mitigation strategy to defend against cybersecurity threads utilizing residual information in the leakage states~\citep{xu2023, tan2023}. 

\section*{Acknowledgments}
We thank Shruti Puri, Steven Girvin, and Aravind Babu for useful discussions. We acknowledge funding by Kvantum Institute of the University of Oulu, and the Research Council of Finland under Grants Nos. 316619, 346035, H2Future Profi7 352788, and 355824. This publication is part of the research and applied innovation project Quantum mechanics and biology: new theoretical frameworks and analytictechnical applications PPIT2024-31572, co-financed by the “UE - Ministerio de Hacienda y Función Pública - Fondos Europeos - Junta de Andalucía – Consejería de Universidad, Investigación e Innovación”. The authors wish  to acknowledge CSC -- IT Center for Science, Finland, for computational resources.\nocite{zenodo}

\appendix
\section{Numerical simulation details}\label{app:numerical_simulation}

The unitary time-evolution of the transmon array is given by the unitary time-evolution operator $\hat{U}=e^{-i\hat{H}dt/\hbar}$, where $dt$ is the time-step. The operator $\hat{U}$ can be solved exactly by diagonalizing the Hamiltonian. This is feasible for small systems, and this method was used for array lengths $L \leq 3$. For larger systems the Krylov-method~\cite{saad_iterative_2003} was used.

The time in between periodic measurements is given by $t_i - t_{i - 1} = 1/\Gamma$, where $\Gamma$ is the periodic measurement rate. The time of the first measurement $t_0$ is chosen randomly. This is important since the time the leakage occurred is not known. The periodic feedback measurement is applied by first performing a measurement of the number operator $\hat{n}$ at the last site. The Born probabilities of different measurement outcomes are calculated. A measurement outcome is randomly chosen based on the calculated probabilities, and the state of the system is projected to match the measurement outcome. Feedback is implemented by operating with a feedback operator $\hat{F}$ on the system following the measurement. Engineer dissipation is implemented by considering the transmon array as an open quantum system, with the jump operator $\sqrt{\Gamma_{\rm d}}\hat{a}_{L}$ acting at the end of the array.

The disorder $W$ of $U_\ell \sim \bar U+[-W,W]$ is implemented by first calculating the on-site energy $\omega_\ell$. We generate a disorder realization by drawing random numbers for each site $\delta \omega_\ell = [-W/2,W/2]$, from which we get the on-site energy realization $\omega_\ell = \bar{\omega} + \delta\omega_\ell$. The on-site energy realization determines the on-site anharmonicity, together with the mean on-site energy $\bar{\omega}$ and the mean on-site anharmonicity $\bar{U}$,
\begin{equation}
    U_\ell = 2\omega_\ell - (2 \bar{w} - \bar{U}) \xrightarrow{} E_2 = 2\bar{w} - \bar{U}.
    \label{parameters_condition}
\end{equation}
This fixes the energy $E_2$ of the second excited level on each site, while having disorder $W/2$ in the first excited energy and disorder $W$ in the anharmonicity.

We use the quantum trajectory approach to simulate open quantum systems \cite{daley14}. For dissipation we use the jump operators $\sqrt{\gamma}\hat{a}_\ell$, where the dissipation rate follows from the relaxation time $\gamma = 1 / T_1^\textrm{q}$. For dephasing we use $\sqrt{2\kappa}\hat{n}_\ell$, where the dephasing rate follows from the pure dephasing time $\kappa = 1 / T_\phi^\textrm{q}$. The factor of two is the result of a transformation from qubit dephasing with the jump operator $\sqrt{\kappa}\hat{\sigma}_z$ to transmon dephasing originating from the number operator $\hat{n}$.

\begin{figure}
\includegraphics[width=1.0\linewidth]{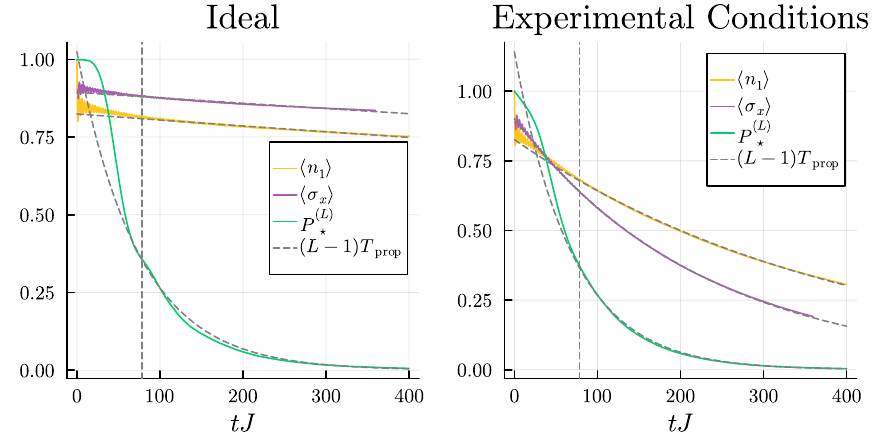}
\caption{\label{fig:fig3_fits}
    Underlying data from which Fig.~\ref{fig:leakage_qubit_effect_realistic} is calculated from. As a function of time: the array leakage population $P_\star^{(L)}(t) = \braket{\sum_{\ell=1}^L \hat{n}_\ell(\hat{n}_\ell-\hat{I})/2}$ with initial state $\ket{\psi_0} = \ket{2}$, the coding qubit excitation number $\braket{\hat{n}_1}$ with initial state $\ket{\psi_0} = \ket{1}$, and the envelope of the oscillations of $\braket{\hat{\sigma}_x}$ with initial state $\ket{\psi_0} = 1/\sqrt{2}(\ket{0} + \ket{1})$. An exponential fit is done to find the constant $\tau$ in the exponent $e^{-t/\tau}$. These constants are then shown in Fig.~\ref{fig:leakage_qubit_effect_realistic}. The figures here show the data for the ideal and experimentally relevant conditions of Fig.~\ref{fig:leakage_qubit_effect_realistic}(a) and Fig.~\ref{fig:leakage_qubit_effect_realistic}(d) respectively, for disorder $W/2\pi =$~\SI{100}{\mega\hertz}.
} 
\end{figure}

\begin{figure*}
\includegraphics[width=1.0\linewidth]{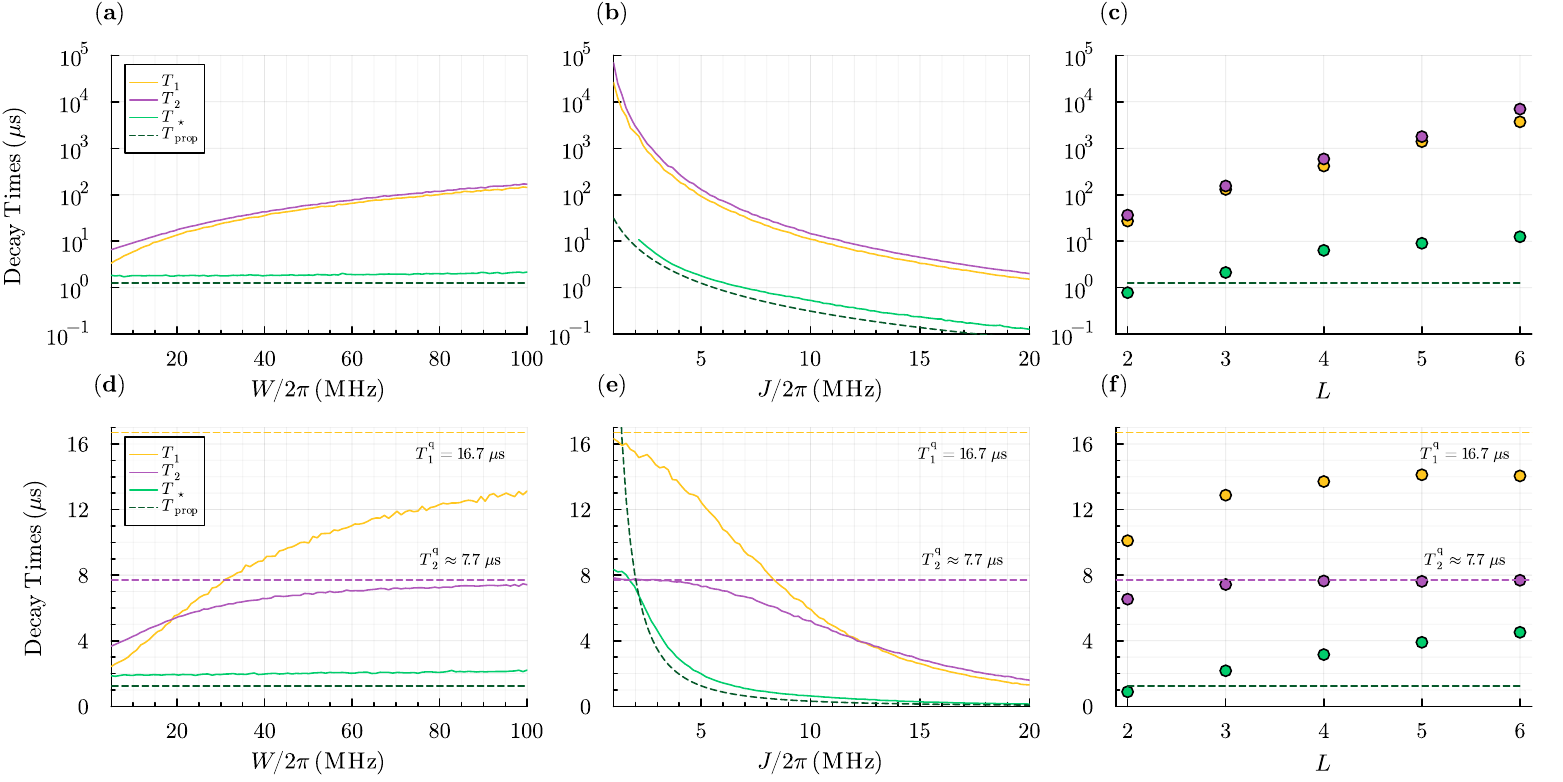}
\caption{\label{fig:decay_diss_low}
    LRU performance for the low optimal dissipation rate $\Gamma_{\rm{d}}^{\rm{low}}$: leakage population decay time $T_{\star}$, leakage population propagation time~$T_{\rm prop}$, qubit dissipation time~$T_1$, and qubit decoherence time~$T_2$ under ideal [ panels (a)-(c)] and experimentally relevant conditions [panels (d)-(f)]. We use the same parameters as in Fig.~\ref{fig:leakage_qubit_effect_realistic}, but with fewer trajectories (4000), and with $\Gamma_{\rm{d}}^{\rm{low}} = 0.04J =2\pi\times$~\SI{0.2}{\mega\hertz}.
} 
\end{figure*}

To take into account non-zero thermal environments, we assume that the initial state of the transmon array excluding the coding qubit is given by the Gibbs state $\hat{\rho} = e^{-\beta\hat{H}} / \text{Tr}(e^{-\beta\hat{H}})$. We use the Bose-Hubbard Hamiltonian with $J = 0$. The resulting density operator is diagonal $\hat{\rho} = \sum p_i \ket{E_i}\bra{E_i}$. The probabilities $p_i$ are used to sample a many-body eigenstate $\ket{E_i}$ of the Bose-Hubbard Hamiltonian. In each trajectory we sample a new many-body eigenstate. The initial state of the transmon array is then given by the tensor product of the initial state of the coding qubit, and the sampled many-body eigenstate.

To calculate the decay times in Fig.~\ref{fig:leakage_qubit_effect_realistic}, we assume that the decays of the array leakage population, qubit population and qubit coherence are exponential. 
An example of the exponential fitting is shown in more detail in Fig.~\ref{fig:fig3_fits}. The optimal measurement rate is sensitive to the nearest-neighbor hopping rate $J$, and to the array length $L$. For the nearest-neighbor hopping rate analysis, in Fig.~\ref{fig:leakage_qubit_effect_realistic}(b, e), we used the analytically derived relation for the optimal measurement rate $\Gamma_{\rm fb}^{\rm{low}}=c J_{\rm prop}$, where the constant $c$ is chosen so that for $J/2\pi=$\SI{5}{\mega\hertz} we have $\Gamma_{\rm fb}^{\rm{low}}=0.03J$. The array length dependence was not found to be significant for the numerical simulations, and constant optimal rate $\Gamma_{\rm fb}^{\rm{low}}=0.03J$ was used. For the results here, Figs.~\ref{fig:decay_diss_high} and~\ref{fig:decay_diss_low}, we used similar reasoning. The low optimal dissipation rate $\Gamma_{\rm d}^{\rm{low}}=0.04J$ was used, following from Fig.~\ref{fig:leakage_qubit_effect_realistic}. For the high optimal dissipation rate, we used the value $\Gamma_{\rm d}^{\rm{high}}=2\overline{U}$, following the analytics.

The overall dynamics is solved by using the quantum trajectory approach. Trajectories are averaged over measurement outcomes of the feedback measurement, disorder realizations, dissipation and decoherence events and the different initial states corresponding the non-zero thermal state. Each trajectory has its own disorder and thermal state realization. The figures have the following number of trajectories: Fig.~\ref{fig:leakage_works}(a)-(b) have \SI{20000} trajectories, Fig.~\ref{fig:leakage_works}(c) has \SI{100000} trajectories; Fig.~\ref{fig:leakage_qubit_effect_realistic}(a)-(c) have \SI{80000} trajectories; Fig.~\ref{fig:leakage_qubit_effect_realistic}(d)-(f) have \SI{100000} trajectories; Fig.~\ref{fig:decay_diss_high} has \SI{4000} trajectories in each case; Fig.~\ref{fig:decay_diss_low} has \SI{40000} trajectories in each case; and Fig.~\ref{fig:rnd_msr_optimal_spots} has \SI{10000} trajectories. The high optimal dissipation rate requires a small time-step $dt$ for accurate simulations.

In Sec.~\ref{numerical_results}, we present the results for the numerical simulations using the high optimal dissipation rate. For low optimal dissipation rate, we show the decay times in Fig.~\ref{fig:decay_diss_low}. We see similar performance in decay times as for periodic feedback measurement.

\section{Alternative reset method: Random feedback measurements}\label{app:alternative_reset_methods}
\begin{figure}
\includegraphics[width=1.0\linewidth]{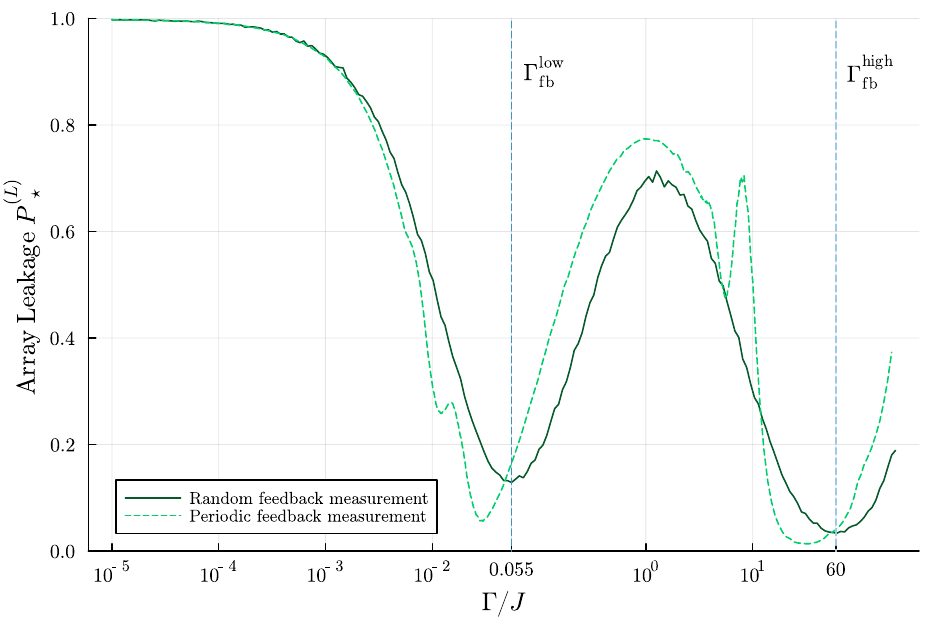}
\caption{\label{fig:rnd_msr_optimal_spots}
    Leakage population in the whole transmon array as a function of the random feedback measurement rates. The leakage population is calculated at time $tJ=200$. Vertical blue dashed lines indicate the two optimal measurement rates $\Gamma_{\rm{fb}}^{\rm{low}}=0.055J$ and $\Gamma_{\rm{fb}}^{\rm{high}}=60J$. The parameters are the same as in Fig.~\ref{fig:leakage_works}.
}
\end{figure}

We present here an alternative way to remove leakage at the last site by using feedback measurements, namely random feedback measurements. We implement random measurements by performing a feedback measurement in each time-step with the probability $p = \Gamma_{\rm fb}dt$. As for the other reset methods presented above, there are two optimal measurement rates, see Fig.~\ref{fig:rnd_msr_optimal_spots}.  We expect the performance to be similar to periodic feedback measurements, and we do not present decay times for random feedback measurement. As it can be seen in Fig.~\ref{fig:rnd_msr_optimal_spots}, we find a similar performance as for periodic feedback measurement. There is a slight change in the optimal measurement rates, but this does not affect the practicality of the device. For low and high optimal rates, there is less leakage removed in the same time when comparing to the periodic feedback measurements. 

\section{Analytical model details}
\label{app:analytical_model}
For the unitary dynamics, we consider an array of $L$ transmons evolving under the attractive Bose-Hubbard Hamiltonian
\begin{equation}
\frac{\hat{H}_{\rm{BH}}}{\hbar} = \sum_{\ell=1}^L \left[\omega_{\ell} \hat{n}_\ell- \frac{U_{\ell}}{2} \hat{n}_{\ell} (\hat{n}_{\ell}-\hat{I})+J_{\ell} \left(\hat{a}_{\ell}^\dagger \hat{a}^{}_{\ell+1}+\textrm{h.c.} \right)\right],
\label{Hamiltonian_BH_app}
\end{equation}
where $\hat{a}_\ell$ and $\hat{a}^\dagger_\ell$ represent the bosonic annihilation and creation operators at site $\ell$, respectively, while $\hat{n}_\ell=\hat{a}^\dagger_\ell \hat{a}^{}_\ell$ denotes the corresponding boson number operator. The term $\omega_\ell$ denotes the on-site energy, and $U_\ell$ represents the attractive interaction strength at site $\ell$ affecting the bosonic excitations. The term~$J_\ell$ indicates the hopping rate of excitations between sites $\ell$ and $\ell+1$, with $J_L$ implicitly accounting for the array's boundary conditions, whether open or periodic. For the LRU, we assume $J_\ell \equiv J$ is constant, and $U_\ell = \bar U+\delta U_\ell$ and $\omega_\ell = \bar \omega+\delta \omega_\ell$ such that $ 2\omega_\ell-U_\ell$ is constant. We consider also open boundary conditions. The Hamiltonian described in Eq.~\eqref{Hamiltonian_BH_app} conserves the total number of excitations, as indicated by $[\hat{H}_{\rm{BH}},\hat{N}]=0$, where $\hat{N}=\sum_{\ell=1}^{L}\hat{n}_\ell$ is the total number operator. This implies that the dynamics occur within a single subspace (also named as a sector) of a fixed number of excitations when the system is initialized with a definite number of excitations. From now on we will take $\hbar=1$.

In what follows, we consider two types of dynamics depending on the number of excitations involved. \textit{Qubit subspace dynamics} refers to $n=0,1$ excitations, and we use the Hamiltonian of Eq.~\eqref{Hamiltonian_BH_app} to describe its dynamics. \textit{Leakage dynamics} refers to $n=2$ excitations located at the same site which is often referred as a boson stack, or stack. We divide this dynamics: \textit{Leakage disintegration} is the fast dynamics in which one of the excitations move to the nearest site, and \textit{Leakage propagation} is the slow dynamics in which the two excitations move jointly as a whole to the nearest site. We can understand the difference between these two phenomena by considering the simplest case of two transmons $L=2$ in the sector of $N=2$ excitations. In this case, the leakage propagation refers to $\ket{20}\leftrightarrow \ket{02}$ and the leakage disintegration to $\ket{\sfrac{20}{02}} \leftrightarrow \ket{11}$; note that the dynamics occur within the same and different anharmonicity manifolds, respectively. 

We refer to these states as belonging to different or the same anharmonicity manifolds of the Hilbert space, corresponding to eigenstates of the operator $ \frac{1}{2}\hat{n}(\hat{n}-\hat{I})$, with distinct eigenvalues, such that $ \frac{1}{2}\hat{n}(\hat{n}-\hat{I})\ket{\sfrac{20}{02}}=\ket{\sfrac{20}{02}}$ and $\frac{1}{2}\hat{n}(\hat{n}-\hat{I})\ket{11}=0$. Therefore, we only need to study the Hamiltonian
\begin{align}
    \hat{H}_{\textnormal{BH}}-(\omega_1+\omega_2)\hat{I}
    =\begin{pmatrix}
        -\bar U & 0 & \sqrt{2}J \\
        0 &  -\bar U & \sqrt{2}J \\
        \sqrt{2}J & \sqrt{2}J & 0
    \end{pmatrix},
    \label{Hamiltonian_BH_N2}
\end{align}
which is given in the basis $\ket{20}, \ket{02}, \ket{11}$. Note that since the parameters fulfill Eq.~\eqref{parameters_condition} and we can consider $ U_1=\bar U+\delta U$ and $ U_2=\bar U-\delta U$, we get $\bar U = -(\delta \omega_1-\delta \omega_2)+ U_1=(\delta \omega_1-\delta \omega_2)+ U_2 $. We study the time evolution for the initial symmetric state $\ket{\psi(t=0)}=\frac{1}{\sqrt{2}}(1,1,0)^T$, where the components of the density matrix are given by
{\allowdisplaybreaks
\begin{align}    
    \rho_{20,20}&=\rho_{02,02}=\rho_{02,20}=\rho_{20,02} \nonumber \\
    &=\frac{\bar U^2}{4\omega_{\rm dis}^2}\left[ 1-\cos (\omega_{\rm dis} t) \right]+\frac{1}{2} \left[ 1+\cos (\omega_{\rm dis} t) \right] \label{rho_2s_pop}, \\
    \rho_{20,11}&=\rho_{02,11}=\rho_{11,20}^*=\rho_{11,02}^* \nonumber \\
    &=\frac{ \sqrt{2}\bar UJ}{ \omega_{\rm dis}^2} \left[ 1-\cos (\omega_{\rm dis} t) \right]-\frac{i2\sqrt{2}J}{\omega_{\rm dis}}\sin(\omega_{\rm dis} t) \\
    \rho_{11,11}&=\frac{8J^2}{\omega_{\rm dis}^2}\left[ 1-\cos (\omega_{\rm dis} t) \right]. \label{rho_1_pop} 
\end{align}}Then, there are oscillations between the anharmonicity manifolds $\ket{\sfrac{20}{02}}$ and $\ket{11} $ with frequency $\omega_{\rm dis}=\sqrt{\bar U^2+16J^2}$. The populations of the two anharmonicity manifolds evolve as
{\allowdisplaybreaks
\begin{align}
    P_\star^{(L)}(t)&=\frac{\bar U^2}{2\omega_{\rm dis}^2}\left[ 1-\cos (\omega_{\rm dis} t) \right]+\frac{1}{2} \left[ 1+\cos (\omega_{\rm dis} t) \right], \\
    \rho_{11,11} (t)&=\frac{8J^2}{\omega_{\rm dis}^2}\left[ 1-\cos (\omega_{\rm dis} t) \right]. \label{rho_2_1_ent}
\end{align}}
In the case of $\bar U=0$, we have that
{\allowdisplaybreaks
\begin{align}
    P_\star^{(L),\ \bar U=0}(t)&=\frac{1}{2} \left[ 1+\cos (4J t) \right] , \\
    \rho_{11,11}^{\bar U=0}(t)&=\frac{1}{2}\left[ 1-\cos (4J t) \right].
    \label{rho_2_1_U0_ent}
\end{align}}Next, we briefly discuss the parameter values that allow us to determine whether the leakage excitations propagates or disintegrate by examining the populations at $\rho_{11,11} (t=t_{\rm{dis}})$, where $t_{\rm{dis}}$ represents the disintegration time. For example, in the case of Eq.~\eqref{rho_2_1_U0_ent}, we have that $ \rho_{11,11}^{\bar U=0}(t_{\rm{dis}}=\frac{\pi}{4J})=1$, and the concept of leakage propagation is not correct. We can evaluate the threshold of $\bar U$ above which the leakage propagation becomes effective. We find that for $\bar U=4J$, the populations in both subspaces are equal
$ \rho_{11,11}(t_{\rm{dis}}) =P_\star^{(L)}(t_{\rm{dis}}) =\frac{1}{2}$ at $t_{\rm{dis}}=\frac{\pi}{4\sqrt{2}J}$; and, being more restrictive, we find that for $\bar U=4 \sqrt{2} J$ all states are equivalent $ \rho_{11,11}(t_{\rm{dis}})=\rho_{20,20}(t_{\rm{dis}})=\rho_{02,02}(t_{\rm{dis}})=\frac{1}{3}$ at $t_{\rm{dis}}=\frac{\pi}{4\sqrt{3}J}$.

The previous cases were done considering an entangled initial state $\ket{\psi(0)}=\frac{1}{\sqrt{2}}(\ket{20}+\ket{02})$. Now we consider a more intuitive and physically meaningful case for LRU, where the initial state corresponds to a leakage excitation located at the first site, $\ket{\psi(0)}=\ket{20}$, where we find the population time evolution
{\allowdisplaybreaks
\begin{align}
    \rho_{20,20}(t)=&\left[ \frac{\bar U}{2\omega_{\rm dis}} \sin{\left( \frac{\omega_{\rm dis}t}{2} \right)} +\frac{1}{2}  \sin{\left( \frac{\bar Ut}{2} \right)}\right]^2 \nonumber \\
    &+\left[ \frac{1}{2} \cos{\left( \frac{\omega_{\rm dis}t}{2} \right)} +\frac{1}{2}  \cos{\left( \frac{\bar Ut}{2} \right)}\right]^2, \\
    \rho_{02,02}(t)=&\left[ \frac{\bar U}{2\omega_{\rm dis}} \sin{\left( \frac{\omega_{\rm dis}t}{2} \right)} -\frac{1}{2}  \sin{\left( \frac{\bar Ut}{2} \right)}\right]^2 \nonumber \\
    &+\left[ \frac{1}{2} \cos{\left( \frac{\omega_{\rm dis}t}{2} \right)} -\frac{1}{2}  \cos{\left( \frac{\bar Ut}{2} \right)}\right]^2, \\
    \rho_{11,11}(t)=&\frac{8J^2}{\omega_{\rm dis}^2} \sin^2{\left( \frac{\omega_{\rm dis}t}{2}\right)}.
\end{align}}For the $\bar U=0$ case, we find the disintegration time from $ \rho_{11,11}^{\bar U=0}(t_{\rm{dis}}=\frac{\pi}{4J})=1$. The case $\bar U \neq 0$ is less clear, but we can consider the threshold as $ \rho_{11,11}(t_{\rm{dis}}) \approx \rho_{20,20}(t_{\rm{dis}})>\rho_{02,02}(t_{\rm{dis}}) $, which implies that at the time of disintegrating, the probabilities are equivalent between $ \rho_{11,11}$ and $\rho_{20,20} $ and larger than $\rho_{02,02} $, so there is no sense in considering a leakage propagating. We find a good estimate by calculating the values at the time of disintegrating $\rho_{11,11}(t_{\rm{dis}})=\rho_{20,20}(t_{\rm{dis}})$, with
\begin{align}
 \sin{\left( \frac{\bar U\pi}{2\omega_{\rm dis}}  \right)}=\frac{8J^2-\bar U^2}{\bar U\omega_{\rm dis}} \quad \to \quad \bar U \approx 1.8 J.
\end{align}
Therefore, we interpret that for $\bar U > 1.8J$ the picture of a leakage propagating is correct. It is clear that the leakage propagation picture become more accurate as we increase the interaction strength $\bar{U} $~\cite{mansikkamaki22}; in the limit $\bar U \to \infty$ we recover the hardcore bosons model. In this article, we consider values $\bar U /J\sim [12.5, 250]$. Interestingly, up to second order in $J/\bar{U}$, the Bose-Hubbard Hamiltonian of Eq.~\eqref{Hamiltonian_BH_app} can be perturbatively approximated as an effective Hamiltonian describing the leakage propagation, given by Ref.~\citep{mansikkamaki22} as
\begin{equation}
\hat{H}_{\rm{BH}}^{\rm{prop}} = J_{\rm prop} \left[\hat{n}^\alpha_1+\hat{n}^\alpha_L-\sum_{\ell=1}^L (\hat{\alpha}_{\ell}^\dagger \hat{\alpha}^{}_{\ell+1}+\textrm{h.c.} ) \right],
\label{Hamiltonian_BH_props_supp}
\end{equation}
where $\hat{\alpha}_\ell$ and $\hat{\alpha}^\dagger_\ell$ represent the leakage excitation annihilation and creation operators at site $\ell$, respectively, while $\hat{n}^\alpha_\ell=\hat{\alpha}^\dagger_\ell \hat{\alpha}^{}_\ell$ denotes the corresponding leakage excitation number operator, and $J_{\rm prop}=2\frac{J^2}{\bar{U}}$ is the effective hopping term. We estimate the time it takes for a leakage excitation to propagate between adjacent sites as $T_{\rm prop}=\pi/(2J_{\rm prop})$. Note that there is an edge-localization effect, causing slower leakage propagation between sites $\ell=1$ and $\ell=2$, as well as between $\ell=L-1$ and $\ell=L$. We explicitly write $J_{\rm prop}$ in the expressions of subsections~\ref{app:fb_prop} and~\ref{app:d_prop} where we make use of the effective model of Eq.~\eqref{Hamiltonian_BH_prop}, while we keep $J$ and $\bar{U}$ in the expressions of the other subsections.

\section{Random feedback measurements}\label{app:fb_analytics}
As a first LRU strategy to remove leakage excitations from the last site, we consider feedback measurements that involves measuring the system, evaluating the classical outcome, and applying a conditional gate. For the simplicity, in the analytics study we consider random feedback measurements instead of the periodic feedback measurements used in the numerical simulations. Random feedback measurements, occurring at a rate $\Gamma_{\rm fb}$, can be understood as measurements happening with a certain probability $p=\Gamma_{\rm fb} dt$ after a small time interval $dt$, while periodic feedback measurements occur at a fixed frequency $\Gamma_{\rm fb}$. Note that in the numerical simulations, errors appear randomly, providing justification for using random measurements as a model. The optimal rates for both types of measurements are equivalent, with differences explained by harmonic effects in the periodic case, as shown in Fig.~\ref{fig:leakage_works} and Fig.~\ref{fig:rnd_msr_optimal_spots} .

Let us start with the general case by considering a density matrix $\hat{\rho}(t)$ at a specific time $t$ representing the state of the system, which undergoes unitary evolution governed by $\hat{U}(dt)=e^{-i\hat{H}_{\rm{BH}}dt}$ over a small period $dt$. Following this evolution, there exists a probability $p=\Gamma dt$ of measuring a site, where $\Gamma$ represents the measurement rate. These measurements involve applying operators $\hat{P}_{\ell,n}$, fulfilling $\sum_{n=0}^{d-1} \hat{P}_{\ell,n}^\dagger \hat{P}_{\ell,n}=\hat{I}_\ell $, which yield $d$ possible outcomes, with $d$ denoting the local Hilbert space dimension. Considering the probabilities for all measurement combinations, the averaged density matrix after $dt$ is expressed as
\begin{align}
    &\hat{\rho}(t+dt)=(1-\Gamma dt)^L \hat{U}(dt)\hat{\rho}(t) \hat{U}^\dagger(dt) \nonumber \\
    &\qquad+(1-\Gamma dt)^{L-1} \Gamma dt  \sum_{\ell=1}^L \sum_{n=0}^{d-1} \hat{P}_{\ell,n}\hat{U}(dt)\hat{\rho}(t) \hat{U}^\dagger(dt) \hat{P}_{\ell,n}^\dagger \nonumber \\
    &\qquad+(1-\Gamma dt)^{L-2} (\Gamma dt)^2 \cdots  \nonumber \\
    &\phantom{\hat{\rho}(t+dt)}=\hat{\rho}(t)-dt L\Gamma \hat{\rho}(t)-idt \hat{H}_{\rm{BH}} \hat{\rho} (t)+i dt \hat{\rho}(t) \hat{H}_{\rm{BH}} \nonumber \\
    &\phantom{\hat{\rho}(t+dt)}\qquad+dt \Gamma   \sum_{\ell=1}^L \sum_{n=0}^{d-1} \hat{P}^{}_{\ell,n} \hat{\rho}(t) \hat{P}_{\ell,n}^\dagger+\mathcal{O}(dt^2).
    \label{averaged_density_matrix_time_step}
\end{align}
By reorganizing terms and taking $dt \to 0$, we obtain the differential equation
\begin{align}
    \frac{d\hat{\rho}(t)}{dt}=-i&[\hat{H}_{\rm{BH}}, \hat{\rho} (t) ]\nonumber \\
    &- \Gamma\left( L\hat{\rho}(t) 
    -   \sum_{\ell=1}^L \sum_{n=0}^{d-1} \hat{P}^{}_{\ell,n} \hat{\rho}(t) \hat{P}_{\ell,n}^\dagger\right)
    \label{averaged_density_matrix_equation_1}.
\end{align}
We can rewrite this equation as the typical master equation by taking into account that we can express the density matrix as
\begin{equation}
    \hat{\rho}(t)
    =\frac{1}{2} \sum_{n=0}^{d-1} \hat{P}_{\ell,n}^\dagger \hat{P}^{}_{\ell,n} \hat{\rho} (t) +\frac{1}{2} \hat{\rho}(t)\sum_{n=0}^{d-1} \hat{P}_{\ell,n}^\dagger \hat{P}^{}_{\ell,n},
\end{equation}
so we have that
\begin{align}
    \frac{d\hat{\rho}(t)}{dt}=&-i[\hat{H}_{\rm{BH}}, \hat{\rho} (t)] \nonumber \\
    &+ \Gamma  \sum_{\ell=1}^L \sum_{n=0}^{d-1} \left( \hat{P}^{}_{\ell,n} \hat{\rho}(t) \hat{P}_{\ell,n}^\dagger-\frac{1}{2}\{\hat{P}_{\ell,n}^\dagger \hat{P}^{}_{\ell,n}, \hat{\rho}(t) \}\right) \nonumber \\
    \equiv&-i [\hat{H}_{\rm{BH}}, \hat{\rho} (t)] \nonumber \\
    &+  \sum_{i=1}^{Ld} \Gamma \left(\hat{L}_{i} \hat{\rho}(t) \hat{L}_{i}^\dagger-\frac{1}{2}  \left\lbrace \hat{L}_{i}^\dagger \hat{L}_{i},\hat{\rho}(t) \right\rbrace \right),
    \label{Lindblad_master_equation}
\end{align}
where we have defined the Lindblad operators $\hat{L}_i \equiv \hat{P}_{\ell,n}$ for indices $i$ running through all sites $\ell$ and measurement outcomes $n$. We obtain a similar equation to Refs.~\citep{fuji20, yamamoto2023} but generalized to a Bose-Hubbard model with a generic local dimension. Note that we can apply these results to any number of measured site; for the LRU we consider only a measurement at the last site.

The master equation~\eqref{Lindblad_master_equation} can be solved using the quantum trajectories approach. First, we define the effective Hamiltonian
\begin{equation}
    \hat{H}_{\rm{NJ}}=\hat{H}_{\rm{BH}}-\frac{i}{2}\Gamma L \hat{I} =\hat{H}_{\rm{BH}}-\frac{i}{2}\Gamma \sum_{\ell=1}^L \sum_{n=0}^{d-1} \hat{P}_{\ell,n}^\dagger \hat{P}^{}_{\ell,n} ,
    \label{H_NJ}
\end{equation}
so we can rewrite Eq.~\eqref{Lindblad_master_equation} as
\begin{equation}
    \frac{d\hat{\rho}(t)}{dt}=-i\hat{H}_{\rm{NJ}} \hat{\rho} (t)+i \hat{\rho}(t) \hat{H}_{\rm{NJ}}^\dagger+\sum_{i=1}^{Ld} \Gamma \hat{L}_{i} \hat{\rho}(t) \hat{L}_{i}^\dagger.    \label{Lindblad_master_equation_effective_hamiltonian}
\end{equation}
This expression of the dynamics allows us to apply the quantum trajectory approach in which Eq.~\eqref{H_NJ} describes the no-jump non-unitary dynamics interrupted by jumps mediated by jump operators $\sqrt{\Gamma}\hat{P}_{\ell,n}$. We consider that the system evolves without jumps from a pure state $\ket{\psi(0)}$ to $\ket{\psi_{NJ}(t)} $ for a period of time $t$
\begin{equation}
    \ket{\psi_{\rm{NJ}}(t)}=e^{-i \hat{H}_{\rm{NJ}} t} \ket{\psi(0)} 
    =e^{-\frac{L\Gamma}{2}t}e^{-i\hat{H}_{\rm{BH}} t} \ket{\psi(0)}.
    \label{no_jump_state}
\end{equation}
Note that the norm of the state $\ket{\psi_{\rm{NJ}}(t)} $ decay with time as
\begin{align}
    \mathscr{N}_{\psi}(t) \equiv & \braket{\psi_{\rm{NJ}}(t)|\psi_{\rm{NJ}}(t)} \nonumber \\
    =&  \bra{\psi(0)}e^{+i \hat{H}_{\rm{NJ}}^\dagger t} e^{-i \hat{H}_{\rm{NJ}} t} \ket{\psi(0)}= e^{-L\Gamma t},
    \label{fb_probability}
\end{align}
After evolving the system for a time $t$, there exists a probability $\Gamma t$ for a jump event to occur, leading to a projection onto any of the system's states
\begin{align}
    \ket{\psi_{\rm{QJ}}(t)}&=\frac{\sqrt{\Gamma} \hat{P}_{\ell,n}\ket{\psi_{\rm{NJ}}(t)}}{\sqrt{\braket{\psi_{\rm{NJ}}(t)|\Gamma \hat{P}_{\ell,n}^\dagger \hat{P}^{}_{\ell,n} |\psi_{\rm{NJ}}(t)}}} \nonumber \\
    &=\frac{\hat{P}_{\ell,n}\ket{\psi_{\rm{NJ}}(t)}}{ ||\hat{P}_{\ell,n} \ket{\psi_{\rm{NJ}}(t)}||},
\end{align}
and the probabilities of projecting to the different states are given by
\begin{equation}
    p_i(t) \equiv p_{\ell,n}(t)=\frac{\Gamma||\hat{P}_{\ell,n} \ket{\psi_{\rm{NJ}}(t)}||^2}{\sum_{\ell'=1}^{L}\sum_{n'=0}^{d-1}\Gamma||\hat{P}_{\ell',n'} \ket{\psi_{\rm{NJ}}(t)}||^2}.
    \label{quantum_trajectories_prob}
\end{equation}
Then, the probability of applying the different operators $\hat{L_i} \equiv \hat{P}_{\ell,n}$ depends solely on Born's rule probability of the projected state. Consequently, by comprehending the unitary dynamics, we can assess the impact of the measurements on the system. For the LRU purposes, we choose feedback measurements that set the outcomes of all measurements to the ground state, i.e.~$\hat{P}_n=\ket{0}\bra{n}$, but the previous analyses is general enough to consider also standard measurements. Finally, it can be proved that by taking averages of $ \ket{\psi_{\rm{NJ}/\rm{QJ}}(t)}\bra{\psi_{\rm{NJ}/\rm{QJ}}(t)} $ over the trajectories we obtain the solution to Eq.~\eqref{Lindblad_master_equation} \citep{daley14}.

\subsection{Leakage propagation}\label{app:fb_prop}
In this subsection, we study the interaction of the random feedback measurements at the second site with the leakage propagation, i.e.~$\ket{20}\leftrightarrow \ket{02}$, so we will consider the model described in Eq.~\eqref{Hamiltonian_BH_prop} for $L=2$. In this subspace, the feedback measurements operators at the second site are given by
\begin{align}
\hat{P}_{2,0}&=\hat{I} \otimes \ket{0} \bra{0} \to \hat{P}_{2,0}= \ket{00}\bra{00}+\ket{20}\bra{20}, \nonumber \\ 
\hat{P}_{2,2}&=\hat{I} \otimes \ket{0} \bra{2} \to \hat{P}_{2,2}=\ket{00}\bra{02},
\end{align}
and we do not consider $\hat{P}_{2,1} $ since the Hamiltonian of Eq.~\eqref{Hamiltonian_BH_prop} describes the dynamic between states $\ket{0}$ and $\ket{2}$, i.e.~no leakage excitation disintegrating. The master equation~\eqref{Lindblad_master_equation} is then
\begin{align}
    \frac{d\hat{\rho}(t)}{dt}=&-i[\hat{H}_{\rm{BH}}^{\rm prop}, \hat{\rho} (t)]  \\
    &+ \Gamma_{\rm fb}  \left(\hat{\rho}(t)- \hat{P}_{2,0} \hat{\rho}(t) \hat{P}_{2,0}^\dagger-\hat{P}_{2,2} \hat{\rho}(t) \hat{P}_{2,2}^\dagger \right), \nonumber
\end{align}
from which we get a set of differential equations
\begin{align}
    \frac{d}{dt}
    \begin{pmatrix}
    \rho_{20,20},
    \rho_{02,02},
    \rho_-,
    \rho_+
    \end{pmatrix}^T
    =
    H
    \begin{pmatrix}
    \rho_{20,20},
    \rho_{02,02},
    \rho_-,
    \rho_+
    \end{pmatrix}^T,
    \label{stack_eq_system_fb}
\end{align}
where the matrix
\begin{equation}
    H=    \begin{pmatrix}
    0 & 0 & i2J_{\rm prop} & 0\\
    0 & -\Gamma_{\rm fb} &  -i2 J_{\rm prop} & 0 \\
    iJ_{\rm prop} & -iJ_{\rm prop} & -\Gamma_{\rm fb} & 0 \\
    0 & 0 & 0 &-\Gamma_{\rm fb}
    \end{pmatrix},
\end{equation}
$\rho_+(t) \equiv \frac{1}{2}(\rho_{02,20}(t)+\rho_{20,02}(t)) $, and  $\rho_-(t) \equiv \frac{1}{2}(\rho_{02,20}(t)-\rho_{20,02}(t)) $. The leakage population in the transmon array is given by $P_\star^{(L)}(t) \equiv \rho_{20,20}(t)+\rho_{02,02}(t)$. The set of equations~\eqref{stack_eq_system_fb} has the form $d/dt \ket{x}=\lambda\hat{H}\ket{x}=(\hat{H}_0+\lambda\hat{H}^{'})\ket{x}$; we solve them perturbatively for different values of the perturbation parameter $\lambda$.  See the Appendix of Ref.~\citep{martinvazquez23} for more details on non-Hermitian perturbation theory. For $\Gamma_{\rm fb}/ J_{\rm prop} \ll 1$, we obtain the right and left eigenvectors and eigenvalues
{\allowdisplaybreaks
\begin{align}
    \ket{x_1^0}&=(1, 1, 0, 0)^T,  & E_1^{(0)} & =0, \\ 
    \ket{x_2^0}&=(-1, 1 , 1 , 0)^T, & E_2^{(0)}&=-4i, \\
    \ket{x_3^0}&=(1 , -1 , 1 , 0)^T, & E_3^{(0)} &=4i, \\
    \ket{x_4^0}&=(0 , 0 , 0 , 1)^T, & E_4^{(0)}&=0, \\
    \ket{\tilde{x}_1^0}&=(1 , 1  , 0 ,0)^T, &\tilde{E}_1^{(0)}&=0, \\
    \ket{\tilde{x}_2^0}&=(-1/2 ,1/2  , 1 , 0 )^T, &\tilde{E}_2^{(0)}&=4i, \\
    \ket{\tilde{x}_3^0}&=(1/2 , -1/2  , 1 , 0)^T, &\tilde{E}_3^{(0)}&=-4i, \\
    \ket{\tilde{x}_4^0}&=(0, 0, 0, 1)^T, &\tilde{E}_4^{(0)}&=0.
\end{align}}
The first order correction to the energies are given by 
\begin{align}
    E_1^{(1)}&=-\frac{\Gamma_{\rm fb}}{J_{\rm prop}},  & 
    E_2^{(1)}&=-\frac{3  \Gamma_{\rm fb}}{2J_{\rm prop}}, \nonumber \\
    E_3^{(1)}&=-\frac{3  \Gamma_{\rm fb}}{2J_{\rm prop}}, & 
    E_4^{(1)}&=-\frac{2 \Gamma_{\rm fb}}{J_{\rm prop}}.
\end{align}
The time evolution for an initial leakage excitation in the first site $ \rho_{20,20}(t=0)=1$ and the leakage population decay time are given by
\begin{align}
    \begin{pmatrix}
    \rho_{20,20}(t) \\
    \rho_{02,02}(t)  \\
    \rho_-(t) \\
    \rho_+(t)
    \end{pmatrix}
    =&
    \sum_{n=1}^4 e^{E_n t} \frac{\ket{x_n^0} \bra{\tilde{x}_n^0}}{\braket{\tilde{x}_n^0 |x_n^0} }
    \begin{pmatrix}
    1 \\
    0  \\
    0 \\
    0
    \end{pmatrix}
    \approx
    \frac{1}{2}
    \begin{pmatrix}
    1 \\
    1  \\
    0 \\
    0
    \end{pmatrix}
    e^{-\frac{\Gamma_{\rm fb}}{2}t} \nonumber \\
    & \to \
    P_{\star,\rm low}^{(L)}(t) \approx e^{-\frac{\Gamma_{\rm fb}}{2}t}.
    \label{lekage_decays_stack_lim1}
\end{align}
For $ J_{\rm prop}/ \Gamma_{\rm fb} \ll 1$, we obtain the eigenvectors and eigenvalues
{\allowdisplaybreaks
\begin{align}
    \ket{x_1^0}&=(1 , 0  , 0 ,0)^T, & E_1^{(0)}&=0, \\ 
    \ket{x_2^0}&=(0 , 1  , 0 ,0)^T, & E_2^{(0)}&=-1, \\
    \ket{x_3^0}&=(0 , 0  , 1 ,0)^T, & E_3^{(0)}&=-1, \\
    \ket{x_4^0}&=(0 , 0  , 0 ,1)^T, & E_4^{(0)}&=-1.
\end{align}}The second order correction to the energy to the first eigenvector is given by
\begin{equation}
    E_1^{(2)}=-\frac{2J_{\rm prop}^2}{ \Gamma_{\rm fb}^2}.
\end{equation}
The time evolution for an initial leakage excitation in the first site $ \rho_{20,20}(t=0)=1$ and the leakage population decay time are given by
\begin{equation}
    \begin{pmatrix}
    \rho_{20,20}(t) \\
    \rho_{02,02}(t)  \\
    \rho_-(t) \\
    \rho_+(t)
    \end{pmatrix}
     \approx
    \begin{pmatrix}
    1 \\
    0  \\
    0 \\
    0
    \end{pmatrix}
    e^{-\frac{2J_{\rm prop}^2}{\Gamma_{\rm fb}}t}
    \ \to \
    P_{\star,\rm low}^{(L)}(t) \approx e^{-\frac{2J_{\rm prop}^2}{ \Gamma_{\rm fb}}t}.
    \label{lekage_decays_stack_lim2}
\end{equation}
Since we have the expressions for the corrected energy in the limits of large, Eq.~\eqref{lekage_decays_stack_lim2}, and small, Eq.~\eqref{lekage_decays_stack_lim1}, measurement rates, we can make a rough estimation for an energy function $ E_1^\star$ for the whole range of measurement rates
\begin{align}
    E_1 \approx&
    \begin{cases}
    -\Gamma_{\rm fb}/2, \quad & \Gamma_{\rm fb}/J_{\rm prop} \ll 1 \\
    -2J_{\rm prop}^2/ \Gamma_{\rm fb}, \quad &J_{\rm prop}/\Gamma_{\rm fb} \ll 1
    \end{cases} \nonumber \\
    \to
    E_1^\star \approx&-\frac{2J_{\rm prop}\Gamma_{\rm fb}}{4J_{\rm prop}^2+ \Gamma_{\rm fb }^2}.
\end{align}
We can approximate the time evolution of the leakage population for the whole range of $\Gamma_{\rm fb}$ as
\begin{equation}
    P_{\star,\rm low}^{(L)}(t) \approx \exp\left(-\frac{t}{T_\star^{\rm{low},\ \rm{fb}}}\right)=\exp{\left(-\frac{ 2J_{\rm prop}^2\Gamma_{\rm fb}}{4J_{\rm prop}^2+ \Gamma_{\rm fb}^2 }t \right)},
    \label{leakage_fb_propagation}
\end{equation}
so that the low optimal rate is $\Gamma^{\rm{low}}_{\rm fb} \approx 2 J_{\rm prop}$, and the decay time at the low optimal rate yields $T_\star^{\rm{low},\ \rm{fb}}(\Gamma_{\rm fb} = \Gamma_{\rm fb}^{\rm{low}}) \approx 2/J_{\rm prop}$.

\subsection{Leakage disintegration}
In this subsection, we study the interaction of the random feedback measurements at the second site with the leakage disintegration, i.e.~$\ket{\sfrac{20}{02}}\leftrightarrow \ket{11}$. The master equation is then given by
\begin{align}
    \frac{d\hat{\rho}(t)}{dt}=&-i[\hat{H}_{\rm{BH}}, \hat{\rho} (t)]  \label{Lindblad_master_equation_fb_splitting} \\
    &+ \Gamma_{\rm fb}  \sum_{n=0}^{2} \left( \hat{P}_{2,n} \hat{\rho}(t) \hat{P}_{2,n}^\dagger-\frac{1}{2}\{\hat{P}_{2,n}^\dagger \hat{P}_{2,n}, \hat{\rho}(t) \}\right). \nonumber
\end{align}
The direct analytical solution in this case is quite complex, so we consider a different approach. Since the effects of feedback measurement can be analyzed by studying combination of the quantum state evolution and measurement backactions through the quantum trajectory picture, see Eq.~\eqref{quantum_trajectories_prob}, we consider standard measurements $ \hat{P}_{n}=\ket{n}\bra{n}$ and study the time dynamics at large times to reach the steady state. Therefore, we consider a measurement rate for a random standard measurement $\Gamma_{\rm st}$. Since the Lindblad master equation~\eqref{Lindblad_master_equation} is linear in $\hat{\rho}$, we can write $\frac{\partial \hat{\rho}(t)}{\partial \tau}=\mathcal{L}\hat{\rho}(t)$, where $\mathcal{L}$ is the Liouvillian superoperator, which preserves the trace and generates a completely positive map $ e^{\mathcal{L}t}$ that describes the time evolution of the system. We follow the scheme neatly described in Ref.~\citep{macieszczak2016, mingatti2018}. We linearize the density matrix and obtain
\begin{align}
    \Bar{\Bar{\mathcal{L}}}\vec{\rho}(t)=&\left[-i(\hat{H}_{\rm{BH}} \otimes \hat{I}-\hat{I} \otimes \hat{H}_{\rm{BH}}^T)  \right. \nonumber \\ 
    &+\Gamma  \sum_{\ell=1}^L \sum_{n=0}^{d-1} \left( \hat{P}_{\ell,n} \otimes \hat{P}_{\ell,n}^* -\frac{1}{2}\hat{P}_{\ell,n}^\dagger \hat{P}^{}_{\ell,n} \otimes \hat{I} \right. \nonumber \\
    &\qquad \qquad \qquad \quad \left. \left.-\frac{1}{2}\hat{I} \otimes \hat{P}_{\ell,n}^T \hat{P}_{\ell,n}^*  \right) \right] \vec{\rho}(t).
    \label{Liouvillian_vectorized}
\end{align}
If we manage to diagonalize $\mathcal{L}$, we can employ the eigenstates as a basis for the Liouvillian space, except for exceptional points \citep{Heiss2012, macieszczak2016}. Consequently, an operator $\hat{A}$ can be uniquely decomposed as $\hat{A}=\sum_i c_i \hat{\rho}_i$, that is, the eigendecomposition. The spectrum always satisfies $\textnormal{Re}[\lambda_i] \leq 0$ \citep{breuer2007, Rivas2012}. The real part of the eigenvalues governs the relaxation to the steady state such that $\hat{\rho}_\textnormal{ss}=\lim_{t \to +\infty} e^{\mathcal{L}t} \hat{\rho}(0)$. Sorting the eigenvalues as $|\textnormal{Re}[\lambda_0]|<|\textnormal{Re}[\lambda_1]|<\cdots<|\textnormal{Re}[\lambda_n]|$, the steady state is defined as $\hat{\rho}_{\textnormal{ss}}=\hat{\rho}_0/\tr[\hat{\rho}_0]$, where $\lambda_0=0$. We can also define the Liouvillian gap (or asymptotic decay rate) \citep{kessler2012} as $\lambda=|\textnormal{Re}[\lambda_1]|$, which describes the slowest relaxation dynamics in the long-time limit. For $\hat{\rho}(t)$ to be physically meaningful, it has to be Hermitian, positive-definite, and with trace one. Considering the existence of a single steady state, and that $\tr[\hat{\rho}(t)]=1$, we have that
\begin{equation}
    \hat{\rho}(t)=\frac{\hat{\rho}_0}{\tr[\hat{\rho}_0]}+\sum_{i \neq 0} c_i(t)\hat{\rho}_i=\hat{\rho}_\textnormal{ss}+\sum_{i \neq 0} c_i(0)e^{\lambda_i t}\hat{\rho}_i,
    \label{Rho_physical}
\end{equation}
where we need to differentiate the cases where $\lambda_i$ are real or complex \citep{macieszczak2016}.

First, we demonstrate the mapping of Rabi oscillations in a driven and non-measured qubit and the oscillations between states $\ket{\sfrac{20}{02}} $ and $\ket{11} $ in two qutrits. This mapping provides us with a simple model to study the significant characteristics of leakage disintegration in transmons and their interaction with measurements. For the qubit case, we consider the simple driven Hamiltonian in rotating wave approximation
\begin{equation}
    \hat{H}_{\rm{QB}}=\Delta \hat{\sigma}_z+\beta \hat{\sigma}_x=\begin{pmatrix} \Delta & \beta \\ \beta & -\Delta \end{pmatrix},
    \label{QB_hamiltonian}
\end{equation}
where $\Delta$ is the drive detuning and $\beta$ accounts for the strength of the applied field.
By diagonalizing Eq.~\eqref{QB_hamiltonian} and setting the initial state as $\ket{\psi(t=0}=\ket{0} $, it is straightforward to obtain the time-evolved density matrix components
{\allowdisplaybreaks
\begin{align}
    \rho_{00}(t)=&\frac{1}{2}
        \left[\frac{\Delta^2}{\omega_{\rm qb}^2}(1-\cos(2\omega_{\rm qb}t))+(1+\cos(2\omega_{\rm qb}t))\right], \nonumber \\
        \rho_{01}(t)=&\frac{1}{2}
        \left[ \frac{\Delta \beta}{E^2}(1-\cos(2\omega_{\rm qb}t))+i\frac{\beta}{E}\sin(2\omega_{\rm qb}t)\right], \nonumber \\
        =&\rho_{01}^*(t) \nonumber \\
        \rho_{11}(t)=&\frac{1}{2}
        \left[\frac{\beta^2}{\omega_{\rm qb}^2}(1-\cos(2\omega_{\rm qb}t))\right],
    \label{QB_density_Db}
\end{align}}where $\omega_{\rm qb}=\sqrt{\Delta^2+\beta^2}$, and the system oscillates with frequency $2\omega_{\rm qb}$. If we have $\Delta=0$, then the density matrix is simply
\begin{align}
    \hat{\rho}^{\Delta=0}(t)=\frac{1}{2}\begin{pmatrix}
        1+\cos(2\beta t) & -i\sin(2\beta t) \\
        i\sin(2\beta t) & 1-\cos(2\beta t)
    \end{pmatrix}.
    \label{QB_density_b}
\end{align}
Comparing with Eq.~\eqref{rho_2s_pop}-\eqref{rho_1_pop}, it is clear that the population dynamics of manifolds exhibit Rabi oscillation for $\Delta=\frac{1}{2}U$ and $\beta=2J$. This finding enables us to explore the simple qubit case for insights into the dynamics of $L=2$ qutrits. 

Second, we study Rabi oscillations of a qubit in the presence of measurements, where the master equation is given by
\begin{align}
    \frac{d\hat{\rho}(t)}{dt}=&-i [\hat{H}_{\rm{QB}}, \hat{\rho} (t) ] \notag \\
    &\quad+  \Gamma_{\rm st}   \sum_{n=0}^{1} \left( \hat{P}_{n}\hat{\rho}(t) \hat{P}_{n}^\dagger-\frac{1}{2}\{\hat{P}_{n}^\dagger \hat{P}_{n},\hat{\rho}(t) \}\right) \notag \\\equiv &  \mathcal{L}\hat{\rho}(t) 
\end{align}
where $\hat{P}_{0}=\ket{0}\bra{0}$ and $\hat{P}_{1}=\ket{1}\bra{1} $. Following Eq.~\eqref{Liouvillian_vectorized}, we obtain the vector and matrix version $\frac{d\vec{\rho}(t)}{dt}=\Bar{\Bar{\mathcal{L}}}\vec{\rho}(t)$,
\begin{align}
    \frac{d}{dt} \begin{pmatrix}
        \rho_{00},\rho_{01} , \rho_{10} , \rho_{11} 
    \end{pmatrix}^T
    =
    \Bar{\Bar{\mathcal{L}}}
    \begin{pmatrix}
        \rho_{00} , \rho_{01},\rho_{10} , \rho_{11} 
    \end{pmatrix}^T,
    \label{QB_Liouvillian_vectorized}
\end{align}
where the Liouvillian matrix 
\begin{align}
        \Bar{\Bar{\mathcal{L}}}
        =
        \begin{pmatrix}
        0 & i\beta & -i\beta & 0 \\
        i\beta & -\Gamma_{\rm st}-i2\Delta  &  0 & -i\beta \\
        -i\beta & 0 & -\Gamma_{\rm st}+i2\Delta & i\beta \\
        0 & -i\beta & i\beta & 0
    \end{pmatrix}.
\end{align}
To illustrate the Liouvillian formalism, we include the simple case of on-resonance ($\beta \neq 0$, $\Delta = 0$) with exact analytical expressions. Diagonalizing the Liouvillian from Eq.~\eqref{QB_Liouvillian_vectorized} we obtain the eigenvalues
{\allowdisplaybreaks
\begin{align}
    E_0&=0, \quad E_1=\frac{-\Gamma_{\rm st}-\sqrt{\Gamma_{\rm st}^2-16 \beta^2}}{2}, \nonumber \\
     E_2&=\frac{-\Gamma_{\rm st}+\sqrt{\Gamma_{\rm st}^2-16 \beta^2}}{2}, \quad E_3=-\Gamma_{\rm st},
\end{align}}
and their corresponding eigenmatrices 
{\allowdisplaybreaks
\begin{align}
    \hat{\rho}_0=&\begin{pmatrix}
        1 & 0 \\ 0 & 1
    \end{pmatrix},
    \label{QB_onresonance_rho0}
    \\
    \hat{\rho}_2=&\begin{pmatrix}
    -1 & -i\frac{\left(\Gamma_{\rm st}-\sqrt{\Gamma_{\rm st}^2-16\beta^2} \right)}{4\beta} \\ i\frac{\left(\Gamma_{\rm st}-\sqrt{\Gamma_{\rm st}^2-16\beta^2} \right)}{4\beta} & 1
    \end{pmatrix},
    \\  
    \hat{\rho}_1=&\begin{pmatrix}
        -1 & -i\frac{\left(\Gamma_{\rm st}+\sqrt{\Gamma_{\rm st}^2-16\beta^2} \right)}{4\beta} \\ i\frac{\left(\Gamma_{\rm st}+\sqrt{\Gamma_{\rm st}^2-16\beta^2} \right)}{4\beta} & 1
    \end{pmatrix},
    \\
    \hat{\rho}_3=&\begin{pmatrix}
        0 & 1 \\ 1 & 0
    \end{pmatrix}.
    \label{QB_onresonance_rho3}
\end{align}}Then, we obtain the density matrix time evolution using Eq.~\eqref{Rho_physical}. If the initial state is $\ket{\psi(0)}=\frac{1}{\sqrt{2}}(\ket{0}+\ket{1})$, we have
\begin{align}
    \hat{\rho}(t)=\frac{1}{2} \hat{\rho}_0+\frac{1}{2} \hat{\rho}_3 e^{E_3t}=\frac{1}{2}\begin{pmatrix} 1 & e^{-\Gamma_{\rm st} t} \\  e^{-\Gamma_{\rm st} t} & 1\end{pmatrix}.
\end{align}
If the initial state is $\ket{\psi(0)}=\ket{0}$, for $\Gamma_{\rm st}<4\beta$ we have complex eigenvalues, and the state is given by
\begin{align}
    \hat{\rho}(t)=&\frac{1}{2}\hat{\rho}_0+\left(-\frac{1}{4} \right)(\hat{\rho}_1 e^{E_1t}+\hat{\rho}_1^\dagger e^{E_1^*t})\nonumber \\
    &+\frac{\Gamma_{\rm st}}{4\sqrt{16\beta^2-\Gamma_{\rm st}^2}} i(\hat{\rho}_1 e^{E_1t}-\hat{\rho}_1^\dagger e^{E_1^*t}),
\end{align}
where the density matrix components are
{\allowdisplaybreaks
\begin{align}
    \rho_{00}(t)=&\frac{1}{2}+\frac{1}{2}e^{-\frac{\Gamma_{\rm st}}{2}t}\left[ \cos{\left( \frac{\sqrt{16\beta^2-\Gamma_{\rm st}^2}}{2}t\right)} \right. \nonumber \\
    &\left.+\frac{\Gamma_{\rm st}}{\sqrt{16\beta^2-\Gamma_{\rm st}^2}}\sin{\left( \frac{\sqrt{16\beta^2-\Gamma_{\rm st}^2}}{2}t\right)} \right], \\
    \rho_{11}(t)=&\frac{1}{2}-\frac{1}{2}e^{-\frac{\Gamma_{\rm st}}{2}t}\left[ \cos{\left( \frac{\sqrt{16\beta^2-\Gamma_{\rm st}^2}}{2}t\right)} \right. \nonumber \\
    &\left.+\frac{\Gamma_{\rm st}}{\sqrt{16\beta^2-\Gamma_{\rm st}^2}}\sin{\left( \frac{\sqrt{16\beta^2-\Gamma_{\rm st}^2}}{2}t\right)} \right].
\end{align}}For $\Gamma_{\rm st}>4\beta$ we have real eigenvalues, and the state is given by
\begin{align}
    \hat{\rho}(t)=&\frac{1}{2}\hat{\rho}_0+\frac{1}{4}\left(\frac{\Gamma_{\rm st}}{\sqrt{\Gamma_{\rm st}^2-16\beta^2}}-1\right) \hat{\rho}_1 e^{E_1t} \nonumber \\
    &+\left(-\frac{1}{4}\right)\left(\frac{\Gamma_{\rm st}}{\sqrt{\Gamma_{\rm st}^2-16\beta^2}}+1\right) \hat{\rho}_2 e^{E_2t},
\end{align}
where the density matrix components are
{\allowdisplaybreaks
\begin{align}
    \rho_{00}(t)=&\frac{1}{2}+\frac{1}{2}e^{-\frac{\Gamma_{\rm st}}{2}t}\left[ \cosh{\left( \frac{\sqrt{\Gamma_{\rm st}^2-16\beta^2}}{2}t\right)} \right. \nonumber \\
    &\left.+\frac{\Gamma_{\rm st}}{\sqrt{\Gamma_{\rm st}^2-16\beta^2}}\sinh{\left( \frac{\sqrt{\Gamma_{\rm st}^2-16\beta^2}}{2}t\right)} \right], \\
    \rho_{11}(t)=&\frac{1}{2}-\frac{1}{2}e^{-\frac{\Gamma_{\rm st}}{2}t}\left[ \cosh{\left( \frac{\sqrt{\Gamma_{\rm st}^2-16\beta^2}}{2}t\right)} \right. \nonumber \\
    &\left.+\frac{\Gamma_{\rm st}}{\sqrt{\Gamma_{\rm st}^2-16\beta^2}}\sinh{\left( \frac{\sqrt{\Gamma_{\rm st}^2-16\beta^2}}{2}t\right)} \right].
\end{align}}In every case, the system dissipates to the steady state as $e^{-\frac{\Gamma_{\rm st}}{2}t} $; for $\Gamma_{\rm st}<4\beta $ there are oscillations while for $ \Gamma_{\rm st}>4\beta$ there are not. For $\Gamma_{\rm st}<4\beta$, the Liouvillian gap (or asymptotic decay rate) consists of the complex conjugates eigenmatrices $\hat{\rho}_{1,2}$, indicating oscillations with random resets. Conversely, for $\Gamma_{\rm st}>4\beta$, the Liouvillian gap reduces to the real eigenmatrix $\hat{\rho}_1$, signifying non-oscillating telegraphic behavior resembling Zeno limit. The transition occurs precisely at the exceptional point $\Gamma_{\rm st}=4\beta$. This can be interpreted by considering that the Rabi oscillation of the unitary dynamics is $2\beta$, thus a measurement rate of $\Gamma_{\rm st}=4\beta$ would occur on average halfway through the Rabi oscillations cycle, effectively disrupting the Rabi oscillations. 

The exact expressions for the general off-resonance case ($\beta \neq 0$, $\Delta \neq 0$) case are intricated and do not provide much intuition about the physics. We will utilize perturbation theory to examine the scenario where $\beta/\Delta \ll 1$, corresponding to the experimental transmons case where $J/\bar U \ll 1$. We will consider $\frac{d\vec{\rho}(t)}{dt}=\Bar{\Bar{\mathcal{L}}}\vec{\rho}(t)$, which can be expressed as
\begin{align}
     \frac{d\ket{\rho(t)}}{dt}=\hat{\mathcal{L}}\ket{\rho(t)}=(\hat{\mathcal{L}}_0+\hat{\mathcal{L}^{'}})\ket{\rho(t)},
\end{align}
where $\hat{\mathcal{L}}$ acts as a Hamiltonian in imaginary time dynamics, and $\hat{\mathcal{L}}_0 $ and $\hat{\mathcal{L}}^{'} $ are the unperturbed Hamiltonian and perturbation to the system, respectively. Since $\hat{\mathcal{L}}$ is complex we need to consider non-Hermitian perturbation theory, see Ref.~\citep{sterheim1972, Brody_2013} and the Appendix C of Ref.~\citep{martinvazquez23}. We express the Hamiltonian in adimensional units $ \Delta(\hat{\mathcal{L}}_0/\Delta+\hat{\mathcal{L}}^{'}/\Delta) \to (\hat{\mathcal{L}}_0/\Delta+\hat{\mathcal{L}}^{'}/\Delta) $, where $\bar{\Gamma}_{\rm st}=\Gamma_{\rm st}/\Delta $ and $\bar{\beta}=\beta/\Delta $. We consider the limit $\bar{\beta}=\beta/\Delta \ll 1 $, where the bi-orthogonal basis for $\hat{\mathcal{L}}_0 $ is simply given by
{\allowdisplaybreaks
\begin{align}
    \ket{\phi_0^{(0)}}&=(1,0,0,0)^T, & E_0^{(0)}&=0, \\
    \ket{\phi_1^{(0)}}&=(0,1,0,0)^T, & E_1^{(0)}&=-\bar{\Gamma}_{\rm st}-2 i, \\
    \ket{\phi_2^{(0)}}&=(0,0,1,0)^T, &E_1^{(0)}&=-\bar{\Gamma}_{\rm st}+2 i, \\
    \ket{\phi_3^{(0)}}&=(0,0,0,1)^T, & E_3^{(0)}&=0,  
\end{align}}where $\ket{\phi_n^{(0)}} \equiv \ket{\tilde \phi_n^{(0)}}$ and $E_n^{(0)}=(\tilde E_n^{(0)})^*$. To avoid degeneracy-related problems we redefine the eigenstates $ \ket{\phi_{\pm}^{(0)}}=(1/\sqrt{2})(\ket{\phi_0^{(0)}} \pm \ket{\phi_3^{(0)}})$ and  $ \ket{\tilde \phi_{\pm}^{(0)}}=(1/\sqrt{2})(\ket{\tilde \phi_0^{(0)}} \pm \ket{\tilde \phi_3^{(0)}})$. The corrections up to the second order in $\bar{\beta}$ to the eigenenergies are given by
{\allowdisplaybreaks
\begin{align}
    E_+& \approx 0, \\
    E_1& \approx -\bar{\Gamma}_{\rm st} \left(1-\frac{2\bar{\beta}^2}{\bar{\Gamma}_{\rm st}^2+4} \right)-i 2\left(1+\frac{2\bar{\beta}^2}{\bar{\Gamma}_{\rm st}^2+4} \right),  \\
    E_2& \approx -\bar{\Gamma}_{\rm st} \left(1-\frac{2\bar{\beta}^2}{\bar{\Gamma}_{\rm st}^2+4} \right)+i 2\left(1+\frac{2\bar{\beta}^2}{\bar{\Gamma}_{\rm st}^2+4} \right), \\ 
    E_-& \approx -\frac{4\bar{\Gamma}_{\rm st} \bar{\beta}^2}{\bar{\Gamma}_{\rm st}^2+4}.
    \label{off_resonance_Em}
\end{align}}Even for the value $ \bar \beta=0.32$ corresponding to the minimum value $\bar U/J=12.5$ considered in the paper, the Liouvillian gap $E_-$ stands out clearly, distinctly separated from the other eigenvalues $E_1$ and $E_2$. However, when increasing $\bar{\beta}$, there will be a value for which the eigenvalues $\textnormal{Re}(E_-)$ and $\textnormal{Re}(E_{1,2})$ cross each other, and above it, there will be a crossing point for a certain $\bar{\Gamma}_{\rm st}$ below which the Liouvillian gap would then correspond to $\hat{\rho}_{1,2}$. 

Although we have proceeded with a perturbative analysis for small $\bar{\beta}$, we can estimate the condition for the crossing point of $\hat{\rho}_-$ with $\hat{\rho}_{1,2}$ as $\bar{\Gamma}_{\rm st}=\sqrt{6\bar{\beta}^2-4}$. Consequently, for $\bar{\beta}<\sqrt{2/3}$ no crossing point exists. Conversely, for $\bar{\beta}>\sqrt{2/3}$, we find that $\textnormal{Re}(E_{1,2}) > \textnormal{Re}(E_{-})$, if $\bar{\Gamma}_{\rm st}<\sqrt{6\bar{\beta}^2-4}$ and $\textnormal{Re}(E_{1,2}) < \textnormal{Re}(E_{-})$ if $\bar{\Gamma}_{\rm st}>\sqrt{6\bar{\beta}^2-4}$. Although we do not present the results here, a more accurate study considering $ \bar \Gamma_{st}$ as a perturbation shows that the level crossing occurs at $\bar \beta=\sqrt2$, such that the conditions for this analysis hold when $ \bar \beta \ll \sqrt2$.

The eigenmatrices up to second order in $\bar{\beta}$ are given by
{\allowdisplaybreaks
\begin{align}
    \hat{\rho}_{+}
    & \approx
    \frac{1}{\sqrt{2}} 
    \begin{pmatrix} 
    1 & 0 \\ 
    0 & 1 
    \end{pmatrix},
    \\
    \hat{\rho}_{1}
    &\approx 
    \begin{pmatrix} 
    \frac{\bar{\beta}}{\bar{\Gamma}_{\rm st}^2+4}(-2 -i \bar{\Gamma}_{\rm st}) & 1-\frac{\bar{\beta}^2}{\bar{\Gamma}_{\rm st}^2+4} \\  
    \frac{\bar{\beta}^2}{\bar{\Gamma}_{\rm st}^2+4}(-1-i\frac{\bar{\Gamma}_{\rm st}}{2}) & \frac{\bar{\beta}}{\bar{\Gamma}_{\rm st}^2+4}(+2 +i\bar{\Gamma}_{\rm st})    
    \end{pmatrix}, 
    \\
    \hat{\rho}_{2}
    &\approx
    \begin{pmatrix} 
    \frac{\bar{\beta}}{\bar{\Gamma}_{\rm st}^2+4}(-2 +i \bar{\Gamma}_{\rm st}) & \frac{\bar{\beta}^2}{\bar{\Gamma}_{\rm st}^2+4}(-1+i\frac{\bar{\Gamma}_{\rm st}}{2}) \\ 
    1-\frac{\bar{\beta}^2}{\bar{\Gamma}_{\rm st}^2+4} & \frac{\bar{\beta}}{\bar{\Gamma}_{\rm st}^2+4}(+2-i \bar{\Gamma}_{\rm st})    
    \end{pmatrix},
    \\
    \hat{\rho}_{-}
    &\approx 
    \begin{pmatrix} 
    \frac{1}{\sqrt{2}}-\frac{\bar{\beta}^2}{\bar{\Gamma}_{\rm st}^2+4}\sqrt{2} & \frac{\bar{\beta}}{\bar{\Gamma}_{\rm st}^2+4}\sqrt{2}(i \bar{\Gamma}_{\rm st}+2) \\ 
    \frac{\bar{\beta}}{\bar{\Gamma}_{\rm st}^2+4}\sqrt{2}(-i \bar{\Gamma}_{\rm st}+2) & -\left(\frac{1}{\sqrt{2}}-\frac{\bar{\beta}^2}{\bar{\Gamma}_{\rm st}^2+4}\sqrt{2} \right)   
    \end{pmatrix},
\end{align}}where we have considered the normalization of the wave function as explained in Ref.~\citep{martinvazquez23}. We can study the time evolution using the eigenmatrices. Since the energies of $\hat{\rho}_{1,2}$ are complex conjugates, we consider hermitian linear combinations as we did above. So the most general state is given by
\begin{align}
    \hat{\rho}(t)\approx& \frac{1}{\tr{\hat{\rho}_+}}\hat{\rho}_++A(\hat{\rho}_1 e^{E_1t}+\hat{\rho}_1^\dagger e^{E_1^*t}) \nonumber \\&+iB(\hat{\rho}_1 e^{E_1t}-\hat{\rho}_1^\dagger e^{E_1^*t})+C\hat{\rho}_- e^{E_-t}.
    \label{off_resonance_time_evolution}
\end{align}
Then we find the coefficients $A$, $B$, and $C$ depending on the initial state. We can exactly compute the coefficients for $\hat{\rho}(0)=\ket{0}\bra{0}$, and by substituting them into~\eqref{off_resonance_time_evolution}, we derive the time dynamics for the population $\rho_{00}(t)$. Given that $ \bar{\beta} \ll 1 $, it can be shown that we can neglect the terms $A$ and $B$ and roughly estimate the dissipation when $\textnormal{Re}(E_{1,2}) \ll \textnormal{Re}(E_{-})$, such that
\begin{align}
    \hat{\rho}_{-} \propto e^{-\frac{4\bar{\Gamma}_{\rm st} \bar{\beta}^2}{\bar{\Gamma}_{\rm st}^2+4}t}  \to \rho_{00}(t) & \sim \frac{1}{2}\left(1 + e^{-\frac{4\bar{\Gamma}_{\rm st} \bar{\beta}^2}{\bar{\Gamma}_{\rm st}^2+4}t}\right).
    \label{liouvillian_eigenmatrix}
\end{align}
Since the Liouvillian gap is quite far from the other eigenvalues in this regime, we can use Eq.~\eqref{liouvillian_eigenmatrix} to study the dynamic to reach the steady state. Note that $\bar{\Gamma}_{\rm st}^{\rm{min}}=2$ minimizes Eq.~\eqref{liouvillian_eigenmatrix}, so it corresponds to the measurement rate that yields the fastest dissipation to the steady state. 

Finally, we need to consider an additional step besides substituting $\Delta=\bar U/2$ and $\beta=2J$ to obtain an expression for the removal of the leakage population $P_\star^{(L)}(t)$. Note that the qubit model is different from the transmons case: there is no true correspondence between measurements in the two models. We have made a mapping between the states $\ket{0}$ and $ \ket{1}$ of the qubit to the subspaces of $ \ket{\sfrac{20}{02}} $ and $\ket{11}$ of the transmons, respectively. Since the subspace spanned by $\ket{20} $ and $\ket{02} $ has two dimensions, it is overrepresented with respect to the qubit state $\ket{0}$, i.e.~there are two projectors for the transmons and one projector for the qubit. We can make the following correspondence between subspace projectors
{\allowdisplaybreaks\begin{align}
    \hat{P}_0&=\ket{0}\bra{0} \to \hat{P}_{20+02}=\frac{1}{2}\left(\ket{20}+\ket{02}\right)\left(\bra{20}+\bra{02}\right), \\
    \hat{P}_1&=\ket{1}\bra{1} \to \hat{P}_{2,1}=\ket{11}\bra{11}.
\end{align}}In the transmon case, for measurements performed at the second site, the following subspace projectors correspondence holds
{\allowdisplaybreaks
\begin{align}
    &\left.
    \begin{matrix}
        \hat{P}_{2,0}=\ket{20}\bra{20} \\
        \hat{P}_{2,2}=\ket{02}\bra{02} 
    \end{matrix}
    \right\}
    \to
    2\hat{P}_{20+02}+\textnormal{off-diag. terms},\\
    &\hat{P}_{2,1}=\ket{11}\bra{11} \to \hat{P}_{2,1}=\ket{11}\bra{11}.
\end{align}}The interpretation of this correspondence is that, in the case of the transmons, the rate for $P_\star^{(L)}(t)  \to \rho_{11,11}(t)$, i.e., exiting the leakage population, is slower by a factor of $1/4$ because the master equation involves two copies of the projectors. Therefore, we solve this issue by setting $t \to 4t$, so that the decay of the leakage population in the transmons is given by
\begin{equation}
    P_{\star.\rm high}^{(L)}(t)  \approx \exp{\left(-\frac{t}{T_\star^{\rm{high}, \ \rm{fb}}}\right)} = \exp{\left(-\frac{4J^2\Gamma_{\rm st}}{\Gamma_{\rm st}^2+U^2}t\right)}.
    \label{fb_freq2_decay2}
\end{equation}
Note that, in this subsection, we have analyzed the long time dynamics in the presence of standard measurements, starting in the anharmonicity manifold $ P_\star^{(L)}$ and populating also the other anharmonicity manifold $\rho_{11,11}$ at the steady state. Therefore, if we focus on $ P_\star^{(L)}(t)$, we can make the equivalence $\Gamma_{\rm st} \to \Gamma_{\rm fb} $ in Eq.~\eqref{fb_freq2_decay2}, since this quantity measures the population remaining the anharmonicity manifold $ P_\star^{(L)}$; in the case of feedback measurements there would be a removal of excitations but it would be not relevant with respect to this quantity. We solved this issue by removing the constant factor $1/2$ from Eq.~\eqref{liouvillian_eigenmatrix}, such that in the steady state $P_\star^{(L)} (t\to \infty)=0 $ in Eq.~\eqref{fb_freq2_decay2} while $ \rho_{00}(t \to \infty)=1/2$ in Eq.~\eqref{liouvillian_eigenmatrix}. Note also that, in the case of feedback measurements, we are neglecting the effect of measuring $ \ket{\sfrac{20}{02}} $ and removing two excitations. Therefore, we have that $\Gamma_{\rm fb}^{\rm{high}}\approx\bar U$, and finally obtain a decay time at the high optimal rate of $T_\star^{\rm{high}, \ \rm{fb}}(\Gamma_{\rm fb}=\Gamma_{\rm{fb}}^{\rm high})\approx J^2/2\bar U$.

\subsection{Qubit subspace dynamics}
In this subsection, we study the effect of feedback measurements in the qubit subspace, i.e., how feedback measurements at the second site affect single excitations  at the first site. We consider the Hamiltonian of Eq.~\eqref{Hamiltonian_BH_app}. In this subspace, the feedback measurements at the second site are expressed as
\begin{align}
\hat{P}_{2,0}=&\hat{I} \otimes \ket{0} \bra{0} \to \hat{P}_{2,0}= \ket{00}\bra{00}+\ket{10}\bra{10}, \\
\hat{P}_{2,1}=&\hat{I} \otimes \ket{0} \bra{1} \to \hat{P}_{2,1}=\ket{00}\bra{01}.
\end{align}
The master equation~\eqref{Lindblad_master_equation} is then given by
\begin{align}
    \frac{d\hat{\rho}(t)}{dt}=-&i[\hat{H}_{\rm BH}, \hat{\rho} (t)] \\
    &+ \Gamma_{\rm fb}  \left(\hat{\rho}(t)- \hat{P}_{2,0} \hat{\rho}(t) \hat{P}_{2,0}^\dagger-\hat{P}_{2,1} \hat{\rho}(t) \hat{P}_{2,1}^\dagger \right), \nonumber
\end{align}
from which we get the set of differential equations
\begin{align}
    \frac{d}{dt}
    \begin{pmatrix}
    \rho_{10,10} ,
    \rho_{01,01},
    \rho_-,
    \rho_+ 
    \end{pmatrix}^T
    =
    H
    \begin{pmatrix}
    \rho_{10,10},
    \rho_{01,01},
    \rho_-,
    \rho_+, 
    \end{pmatrix}^T,
\end{align}
where the matrix
\begin{equation}
    H=
    \begin{pmatrix}
    0 & 0 & -i2J & 0\\
    0 & -\Gamma_{\rm fb} &  i2J & 0  \\
    -iJ & iJ & -\Gamma_{\rm fb} & i(\omega_1-\omega_2) \\
    0 &0 &  i(\omega_1-\omega_2) & -\Gamma_{\rm fb}
    \end{pmatrix}.
\end{equation}
We have defined $\rho_+(t) \equiv \frac{1}{2}(\rho_{01,10}(t)+\rho_{10,01}(t)) $ and  $\rho_-(t) \equiv \frac{1}{2}(\rho_{01,10}(t)-\rho_{10,01}(t)) $.
For the particular case $\omega_1=\omega_2$, where there is no disorder, we can solve this system as we did for the leakage propagation, obtaining $\rho_{10,10}(t)+\rho_{01,01}(t) \approx \exp{\left(-\frac{2\Gamma_{\rm fb} J^2}{\Gamma^2+4J^2}t\right)}$, so that the optimal rate for removing one excitation is $2J$. When there is disorder, we can solve the system perturbatively. Considering $ J/\bar \omega \ll 1$ and $\omega_\ell=\bar \omega+\delta \omega_\ell$ we calculate the eigenvectors and eigenvalues
{\allowdisplaybreaks
\begin{align}
    &\ket{x_1^0}=(1 , 0  , 0 , 0 )^T, \  &E_1^{(0)}=&0, \\ 
&\ket{x_2^0}=(0 , 1  , 0 , 0 )^T,& \ E_2^{(0)}=&-\frac{\Gamma_{\rm fb}}{\bar \omega}, \\
    &\ket{x_3^0}=(0 , 0  , -1 , 1 )^T, \ &E_3^{(0)}=&-\frac{\Gamma_{\rm fb}}{\bar \omega}-i \frac{\delta \omega_{12} }{\bar \omega}, \\
    &\ket{x_4^0}=(0 , 0  , 1 , 1)^T,& \ E_4^{(0)}=&-\frac{\Gamma_{\rm fb}}{\bar \omega}+i \frac{\delta \omega_{12}}{\bar \omega},
\end{align}}where $ \ket{x_n^0} = \ket{\tilde{x}_n^0}$, $E_n^{(0)}=(\tilde{E}_n^{(0)})^*$ and $\delta \omega_{ij} \equiv \delta \omega_i-\delta \omega_j$. The second order correction to the energy to the first eigenvector is given by
\begin{equation}
    E_1^{(2)}=-\frac{J^2}{\bar \omega}\frac{2 \Gamma_{\rm fb}}{\Gamma_{\rm fb}^2+(\delta \omega_1 -\delta \omega_2)^2},
\end{equation}
so that the time evolution for an initial single excitation in the first site $ \rho_{10,10}(t=0)=1$ is given by
\begin{equation}
    \begin{pmatrix}
    \rho_{10,10}(t) \\
    \rho_{01,01}(t)  \\
    \rho_-(t) \\
    \rho_+(t)
    \end{pmatrix}
    \approx
    \begin{pmatrix}
    1 \\
    0  \\
    0 \\
    0
    \end{pmatrix}
    \exp{\left(-\frac{2J^2\Gamma_{\rm fb}}{\Gamma_{\rm fb}^2+(\delta \omega_1 -\delta \omega_2)^2}t\right)},
\end{equation}
and the single excitation population decays as $\rho_{10,10}(t)+\rho_{01,01}(t) \approx\exp{\left(-\frac{2J^2\Gamma_{\rm fb}}{\Gamma_{\rm fb}^2+(\delta \omega_1 -\delta \omega_2)^2}t\right)}$. Now, we can calculate $T_1^{\rm{fb}}$ and $T_2^{\rm{fb}}$. For $T_1^{\rm{fb}}$, we have already the result $\braket{\hat{n}_1(t)}\approx e^{-t/T_1^{\rm{fb}}}$, where $T_1^{\rm{fb}}$ is given by
\begin{equation}
    T_1^{\rm{fb}} \approx \frac{\Gamma_{\rm fb}^2+(\delta \omega_1 -\delta \omega_2)^2 }{2 J^ 2\Gamma_{\rm fb}}.
    \label{T1_fb}
\end{equation}
For $T_2^{\rm{fb}}$, we consider an initial state $ \hat{\rho}_+ =\ket{+}\bra{+}=(1/2)(\ket{10}\bra{10}+\ket{10}\bra{00}+\ket{00}\bra{10}+\ket{00}\bra{00})$, so we need to consider the set of equations
\begin{align}
    \frac{d}{dt}
    (
    \rho_{10,00},
    \rho_{00,10},&
    \rho_{01,00},
    \rho_{00,01}
    )^T
    = \nonumber \\
    &H
    \begin{pmatrix}
    \rho_{10,00},
    \rho_{00,10},
    \rho_{01,00},
    \rho_{00,01}
    \end{pmatrix}^T,
\end{align}
where the matrix
\begin{equation}
    H
    =
    \begin{pmatrix}
    -i\omega_1 & 0 & -iJ & 0\\
    0 & -i\omega_1&  0 & iJ  \\
    -iJ & 0 & -i\omega_2-\Gamma_{\rm fb} & 0 \\
    0 & iJ & 0 & +i\omega_2-\Gamma_{\rm fb}
    \end{pmatrix}.
\end{equation}
In the limit $ J/\bar \omega \ll 1$, we obtain the eigenvectors and eigenvalues
{\allowdisplaybreaks
\begin{align}
    \ket{x_1^0}&=(1 , 0  , 0 , 0 )^T, &  E_1^{(0)}&-i\left( 1 +\frac{\delta \omega_1}{\bar \omega} \right), \\
    \ket{x_2^0}&=(0 , 1  , 0 , 0 )^T, & E_2^{(0)}&=i\left( 1 +\frac{\delta \omega_1}{\bar \omega} \right), \\
    \ket{x_3^0}&=(0 , 0  , -1 , 1 )^T, & E_3^{(0)}&=-i\left( 1 +\frac{\delta \omega_2}{\bar \omega} \right)-\frac{\Gamma_{\rm fb}}{\bar \omega}, \\
    \ket{x_4^0}&=(0 , 0  , 1 , 1)^T, &E_4^{(0)}&=i\left( 1 +\frac{\delta \omega_2}{\bar \omega} \right)-\frac{\Gamma_{\rm fb}}{\bar \omega}.
\end{align}}The second order correction of the energies of the relevant eigenvalues are
{\allowdisplaybreaks
\begin{align}
    E_1^{(2)}= - \frac{J^2}{\bar \omega}\frac{[\Gamma_{\rm fb}+i\left(\delta \omega_1-\delta \omega_2 \right)]}{\Gamma_{\rm fb}^2+\left(\delta \omega_1-\delta \omega_2 \right)^2}, \\
    E_2^{(2)}=- \frac{J^2}{\bar \omega}\frac{\left[ \Gamma_{\rm fb}-i\left(\delta \omega_1-\delta \omega_2 \right)\right]}{\Gamma_{\rm fb}^2+\left(\delta \omega_1-\delta \omega_2 \right)^2} .
\end{align}}Then for the initial state $\ket{+}$, we have that
\begin{align}
    \rho_{10,00}(t)&\approx\frac{1}{2} \exp{\left( -\frac{J^2}{\Gamma_{\rm fb}^2+\delta \omega_{12}^2} \left[ \frac{\Gamma_{\rm fb}}{\bar \omega}+i\frac{\delta \omega_{12}}{\bar \omega}\right]t\right)}.
\end{align}Taking into account that $\rho_{00,00}=1-\rho_{10,10}-\rho_{01,01}$, we finally have that
\begin{equation}
    \braket{\hat{\rho}_+}\approx\frac{1}{2}\left\lbrace1+e^{-\frac{t}{T_2^{\rm{fb}}}}\cos{\left[ \left(\omega_1+\frac{J^2 \delta \omega_{12}}{\Gamma_{\rm fb}^2+\delta \omega_{12}^2} \right)t \right]}\right\rbrace,
\end{equation}
where $T_2^{\rm{fb}}$ is given by
\begin{equation}
    T_2^{\rm{fb}}\approx\frac{\Gamma_{\rm fb}^2+(\delta \omega_1 -\delta \omega_2)^2}{J^2\Gamma_{\rm fb}},
    \label{T2_fb}
\end{equation}
and the optimal rate for removing the qubit subspace population is $\Gamma_{\rm fb}^{[T_1]}=\Gamma_{\rm fb}^{[T_2]}\approx|\omega_1 -\omega_2|$. Therefore, for protecting the qubit subspace, we need to choose a value of $\Gamma_{\rm fb}$ far from this optimal rate.

\section{Engineered dissipation}\label{app:dissipation_analytics}
The second strategy for removing leakage errors from the system is to add an engineered dissipation at the last site, so that the master equation is given by
\begin{align}
    \frac{d\hat{\rho}(t)}{dt}=&-i [\hat{H}_{\rm{BH}}, \hat{\rho} (t)]\nonumber \\
    &\quad+ \Gamma_{\rm d} \left(\hat{a}_{L} \hat{\rho}(t) \hat{a}_{L}^\dagger-\frac{1}{2}  \left\lbrace \hat{a}_{L}^\dagger \hat{a}_{L},\hat{\rho}(t) \right\rbrace \right),
    \label{Lindblad_master_equation_dissipation}
\end{align}
where $\Gamma_{\rm d}$ is the dissipation rate. To analyze the effect of dissipation, we focus on the probability of an event of dissipation to happen in the sense described by the quantum trajectory approach of Eq.~\eqref{quantum_trajectories_prob}, given by
\begin{equation}
    p_{\rm{d}}(t)=1-\braket{\psi_{\rm{NJ}}(t)|\psi_{\rm{NJ}}(t)}.
\end{equation}
Then, studying the reduction of the norm of $\ket{\psi_{\rm{NJ}}}$ in terms of the dissipation rate $\Gamma_{\rm d}$ gives us a proper framework to analyze the optimal parameters. In general, the time evolution of a quantum state under the effective Hamiltonian is given by
\begin{align}
    \ket{\psi_{\rm{NJ}}(t)}=&e^{-i\hat{H}_{\rm{NJ}} t}\ket{\psi_{\rm{NJ}}(0)} \nonumber \\
    =&e^{-i\hat{H}_{\rm{NJ}}t}\sum_{n=1}^{\textnormal{dim}(\mathcal{H})} \frac{\ket{\psi_n}\bra{\tilde{\psi}_n}}{\braket{\tilde{\psi}_n|\psi_n}}\ket{\psi_{\rm{NJ}}(0)} \nonumber \\
    =&\sum_{n=1}^{\textnormal{dim}(\mathcal{H})} e^{-iE_nt}\frac{\braket{\tilde{\psi}_n|\psi_{\rm{NJ}}(0)}}{\braket{\tilde{\psi}_n|\psi_n}}\ket{\psi_n},
\end{align}
and the norm $\mathscr{N}_\psi \equiv \braket{\psi_{\rm{NJ}}(t)|\psi_{\rm{NJ}}(t)}$ by
\begin{align}
    \mathscr{N}_\psi =\sum_{m,n=1}^{\textnormal{dim}(\mathcal{H})}e^{-i(E_n-E_m^*)t}&\frac{\braket{\tilde{\psi}_n|\psi_{\rm{NJ}}(0)}\braket{\psi_{\rm{NJ}}(0)|\tilde{\psi}_m}}{\braket{\tilde{\psi}_n|\psi_n}\braket{\psi_m|\tilde{\psi}_m}} \nonumber \\
    &\times \braket{\psi_m|\psi_n},
    \label{norm_evolution}
\end{align}
where $\hat{H}_{\rm{eff}}\ket{\psi_n}=E_n\ket{\psi_n}  $ and $\hat{H}_{\rm{eff}}^\dagger\ket{\tilde{\psi}_n}=\tilde{E}_n\ket{\tilde{\psi}_n} $. For the case of feedback measurements, we showed in Eq.~\eqref{fb_probability} that the norm decreases gradually with increasing $\Gamma_{\rm d}$ until reaching the Zeno effect due to fixing the initial state.

\subsection{Leakage propagation}\label{app:d_prop}
Proceeding as in the case of feedback measurements, we consider an effective model of leakage propagating moving in a subspace of $ \ket{20}$ and $\ket{02}$ described in Eqs.~\eqref{H_NJ} and~\eqref{Hamiltonian_BH_prop}, such that the effective Hamiltonian in the sense described in Eq.~\eqref{norm_evolution} is 
\begin{equation}
    \hat{H}_{\rm{NJ}}=J_{\rm prop} \left[\hat{n}^\alpha_1+\hat{n}^\alpha_2- (\hat{\alpha}_{1}^\dagger \hat{\alpha}^{}_{2}+\hat{\alpha}_{2}^\dagger \hat{\alpha}^{}_{1} )\right]-i\Gamma_{\rm d} \hat{n}^{\alpha}_2.
    \label{L2_diss_Heff}
\end{equation}
This system can be exactly diagonalized avoiding the exceptional point at $\Gamma_{\rm d}=2 J_{\rm prop}$. For $\Gamma_{\rm d}<2 J_{\rm prop}$, we have that
\begin{align}
    \mathscr{N}_{\psi,\rm low} =& \frac{e^{-\Gamma_{\rm d} t}}{4-\frac{\Gamma^2_{\rm d} }{J^2_{\rm prop}}} \bigg[ 4-\frac{\Gamma^2_{\rm d} }{J^2_{\rm prop}} \cos{\left(t \sqrt{4 J_{\rm prop}^2-\Gamma_{\rm d}^2}\right)}  \nonumber \\
    &+2\frac{\Gamma_{\rm d} }{J_{\rm prop}}\sqrt{1-\frac{\Gamma^2_{\rm d} }{4J^2_{\rm prop}}}\sin{\left( t\sqrt{4 J_{\rm prop}^2-\Gamma_{\rm d}^2}\right)} \bigg],
    \label{L2_diss_prop_exact1}
\end{align}
while for $\Gamma_{\rm d}>2 J_{\rm prop}$, we have that
\begin{align}
    \mathscr{N}_{\psi,\rm low} = \frac{e^{-\Gamma_{\rm d} t}}{\frac{\Gamma^2_{\rm d} }{J^2_{\rm prop}}-4} \bigg[ -4+\frac{\Gamma^2_{\rm d} }{J^2_{\rm prop}} \cosh{\left(t \sqrt{\Gamma_{\rm d}^2-4J_{\rm prop}^2}\right)} \nonumber \\
     +2\frac{\Gamma_{\rm d} }{J_{\rm prop}}\sqrt{\frac{\Gamma^2_{\rm d} }{4J^2_{\rm prop}}-1}\sinh{\left(t \sqrt{\Gamma_{\rm d}^2-4J_{\rm prop}^2}\right)} \bigg].
    \label{L2_diss_prop_exact2}
\end{align}
To have a simpler expression for interpretation and comparison purposes with Eq.~\eqref{leakage_fb_propagation}, we also solve the system perturbatively: for $ J_{\rm prop}/\Gamma_{\rm d} \ll 1$, we obtain $ \mathscr{N}_{\psi,\rm low}= {\rm e}^{-\frac{2J_{\rm prop}^2}{ \Gamma_{\rm d}}t}$, and for $ J_{\rm prop}/ \Gamma_{\rm d} \gg 1$ we obtain $ \mathscr{N}_{\psi,\rm low}= {\rm e}^{-\Gamma_{\rm d} t}$. Therefore, we have the approximate expression $\mathscr{N}_{\psi,\rm low} \approx \exp \left(-\frac{t}{T_\star^{\rm{low}, \ \rm{d}}} \right)=\exp{\left(-\frac{2J_{\rm prop}^2\Gamma_{\rm d}}{2J_{\rm prop}^2+\Gamma_{\rm d}^2}t \right)}$, where we get the optimal rate $\Gamma_{\rm d}^{\rm{low}}=\sqrt{2}J_{\rm prop}$. We get an approximate expression for the decay time for the leakage population at the low optimal rate of $T_\star^{\rm{low}, \ \rm{d}}(\Gamma_{\rm d}=\Gamma_{\rm d}^{\rm{low}})=\sqrt{2}/J_{\rm prop}$.

\subsubsection{$L=3$ with edge-localization}
We can also solve perturbatively the case of $L=3$, where the effective Hamiltonian is given by
\begin{equation}
    \hat{H}_{\rm{NJ}}=J_{\rm prop} \left[\hat{n}^\alpha_1+\hat{n}^\alpha_3-\sum_{\ell=1}^3 (\hat{\alpha}_{\ell}^\dagger \hat{\alpha}^{}_{\ell+1}+\textrm{h.c.} ) \right]-i\Gamma_{\rm d} \hat{n}^{\alpha}_3.
    \label{L3_diss_Heff}
\end{equation}
We solve this system in the limit of $\Gamma_{\rm d} /J_{\rm prop} \ll 1$, where we have that
\begin{align}
    \hat{H}_0
    =&
    \hat{n}^\alpha_1+\hat{n}^\alpha_3-\sum_{\ell=1}^3 (\hat{\alpha}_{\ell}^\dagger \hat{\alpha}^{}_{\ell+1}+\textrm{h.c.} ),
    \notag \\
    \hat{H}^{'}
    =&
    -\frac{i \Gamma_{\rm d} }{J_{\rm prop}}
    \hat{n}^{\alpha}_3.
\end{align}
The unperturbed eigenvectors and eigenvalues are
{\allowdisplaybreaks
\begin{align}
    \ket{\psi_1^0}=&\frac{1}{\sqrt{2}} (-1 , 0 , 1 )^T, & E_1^{(0)} &=1, \\ 
    \ket{\psi_2^0}=&\frac{1}{\sqrt{2}}(1 , 2 , 1 )^T, & E_2^{(0)}&=-1, \\
    \ket{\psi_3^0}=&\frac{1}{\sqrt{2}} (1 , -1 , 1 )^T, & E_3^{(0)}&=2.
\end{align}}
The first order correction is then given by
\begin{align}
    E_1^{(1)} &=-\frac{\Gamma_{\rm d}}{2 J_{\rm prop}}i, &
    E_2^{(1)}&=-\frac{\Gamma_{\rm d} }{6 J_{\rm prop}}i,  \notag \\
    E_3^{(1)} &=-\frac{\Gamma_{\rm d} }{3 J_{\rm prop}}i.
\end{align}
We solve this system in the limit of $J_{\rm prop}/\Gamma_{\rm d}  \ll 1$, where we have that
\begin{align}
    \hat{H}_0
    =&
    -i\hat{n}^{\alpha}_3,
    \\
    \hat{H}^{'}
    =&
    \frac{J_{\rm prop} }{\Gamma_{\rm d} }
    \left[\hat{n}^\alpha_1+\hat{n}^\alpha_3-\sum_{\ell=1}^3 (\hat{\alpha}_{\ell}^\dagger \hat{\alpha}^{}_{\ell+1}+\textrm{h.c.} )\right].
\end{align}
The unperturbed eigenvectors and eigenvalues are
{\allowdisplaybreaks
\begin{align}
    \ket{\psi_1^0}=&(\alpha , 1 , 0 )^T,\quad E_1^{(0)}=0, \\ 
    \ket{\psi_2^0}=&(\beta  , 1 , 0 )^T,\quad E_2^{(0)}=0, \\
    \ket{\psi_3^0}=& (0 , 0 , 1 )^T,\quad E_3^{(0)}=-i,
\end{align}}where $a=-\frac{1}{2}(1-\sqrt{5})$ and $b=-\frac{1}{2}(1+\sqrt{5})$. The first order correction is then given by
\begin{align}
    E_1^{(1)}=&\frac{a^2-2a}{1+a^2}\frac{J_{\rm prop}}{\Gamma_{\rm d} }, & 
    E_2^{(1)}=&\frac{b^2-2b}{1+b^2}\frac{J_{\rm prop}}{\Gamma_{\rm d} },  \nonumber\\
    E_3^{(1)}=&\frac{J_{\rm prop}}{\Gamma_{\rm d} },
    \label{L3_diss_firstcorrection}
\end{align}
which are pure real corrections. The second order corrections are given by
{\allowdisplaybreaks
\begin{align}
    E_1^{(2)}=&-\left( \frac{J_{\rm prop}}{\Gamma_{\rm d} } \right)^2\frac{1}{1+a^2}i,  \\
    E_2^{(2)}=&-\left(  \frac{J_{\rm prop}}{\Gamma_{\rm d} } \right)^2\frac{1}{1+b^2}i,  \\
    E_3^{(3)}=&\left(  \frac{J_{\rm prop}}{\Gamma_{\rm d} } \right)^2 \left[\frac{1}{1+a^2}+\frac{1}{1+b^2} \right]i.
\end{align}}
We finally obtain the norm in the two limits
\begin{align}
    \mathscr{N}_{\psi,\rm low} \approx 
    \begin{cases}
    \frac{1}{2}e^{-\Gamma_{\rm d} t}+\frac{1}{6}e^{-\frac{1}{3}\Gamma_{\rm d} t}+\frac{1}{3}e^{-\frac{2}{3}\Gamma_{\rm d} t}, \ \frac{\Gamma_{\rm d}}{J_{\rm prop}} \ll 1 \\
    \frac{a^2}{(1+a^2)^2}\exp{\left(-\frac{2 J_{\rm prop}^2}{ \Gamma_{\rm d}}\frac{1}{1+a^2}t \right)}& \\+\frac{b^2}{(1+b^2)^2}\exp{\left(-\frac{2J_{\rm prop}^2}{ \Gamma_{\rm d}}\frac{1}{1+b^2}t \right)}, \frac{\Gamma_{\rm d}}{J_{\rm prop}} \gg 1 
    \end{cases}
    \label{L3_diss_prop_edgelocalization}
\end{align}

\subsubsection{General $L$ without edge-localization}
Note in Eqs.~\eqref{L2_diss_Heff} and~\eqref{L3_diss_Heff} that, in the effective model for leakage propagating, the first and last sites acquire different on-site energy compared to the other sites, introducing disorder that affects the overall dynamics \citep{mansikkamaki21}. Although deriving an analytical expression for a generic $L$ is more complex, we can easily obtain one for the case that disregards border effects. This could help us understand the effect of chain length on leakage excitation removal, and it also describes a system where the stack appears and is removed at intermediate sites. Thus, the Hamiltonian
\begin{equation}
    \hat{H}_{\rm{NJ}}=-J_{\rm prop}\sum_{\ell=1}^L (\hat{\alpha}_{\ell}^\dagger \hat{\alpha}^{}_{\ell+1}+\textrm{h.c.} ) -i\Gamma_{\rm d} \hat{n}^{\alpha}_L,
\end{equation}
could be understood as a subspace of a longer chain. For $\Gamma_{\rm d} /J_{\rm prop} \ll 1$, we have that
\begin{equation}
    \hat{H}_{0}
    =
    -\sum_{\ell=1}^L (\hat{\alpha}_{\ell}^\dagger \hat{\alpha}^{}_{\ell+1}+\textrm{h.c.} ),
    \quad
    \hat{H}^{'}
    =
    -\frac{i\Gamma_{\rm d} }{J_{\rm prop}}\hat{n}^{\alpha}_L.
\end{equation}
Switching to the reciprocal space~\cite{mansikkamaki21}, we can express this terms as
\begin{align}
    \hat{H}_{0}=&-2\sum_{k=1}^L \cos{\left( \frac{k\pi}{L+1} \right)} \hat{c}_k^\dagger \hat{c}^{}_k, \\
    \hat{H}^{'}=&\frac{2}{L+1}\left(-\frac{i\Gamma_{\rm d} }{J_{\rm prop}}\right) \nonumber \\ &\times \sum_{j,k=1}^L \sin{\left( \frac{L j\pi}{L+1} \right)} \sin{\left( \frac{L k\pi}{L+1} \right)}\hat{c}_j^\dagger \hat{c}^{}_k.
    \label{L_eigen_reciprocal}
\end{align}
The eigenvectors and eigenvalues of $\hat{H}_{0}$ in the reciprocal space are given by
\begin{align}
    \ket{\psi_\ell^0}&=(0,  0,   \cdots,  1_\ell,  \cdots  0 )^T,\quad E_\ell^{(0)}=-2 \cos{\left( \frac{\ell \pi}{L+1} \right)},
\end{align}
where $1_\ell$ refers to the location in the vector and $\ell=1,2,...,L$. The first order correction are given by
\begin{equation}
    E_\ell^{(1)}=-\frac{2}{L+1}\left( \frac{\Gamma_{\rm d} }{J_{\rm prop}} \right) \sin^2{\left( \frac{\ell L \pi}{L+1} \right)}i.
\end{equation}
To evaluate the initial state in the real space, we have that the zero order eigenvectors of Eq.~\eqref{L_eigen_reciprocal} in the real space are given by
\begin{align}
    \ket{\psi_\ell^0}&=\frac{1}{\sqrt{\mathscr{N}_\ell}}\left( \sin{\left( \frac{\ell\pi}{L+1} \right)}, \sin{\left( \frac{2\ell \pi}{L+1} \right)},   \cdots   \right. \nonumber \\
    &\quad\qquad\qquad \left. ,\sin{\left( \frac{\ell^2 \pi}{L+1} \right)},  \cdots,\sin{\left( \frac{L\ell \pi}{L+1} \right)} \right)^T, \nonumber \\ \mathscr{N}_\ell&=\sum_{k=1}^L \sin^2{\left( \frac{k\ell \pi}{L+1} \right)},
\end{align}
where
\begin{align}
    \mathscr{N}_\psi \approx \sum_{\ell=1}^L& \frac{\sin^2{\left( \frac{\ell \pi}{L+1} \right)} }{\sum_{k=1}^L \sin^2{\left( \frac{k\ell \pi}{L+1} \right)}} \nonumber \\
    &\times\exp{\left[-\frac{4}{L+1}\sin^2{\left( \frac{\ell L \pi}{L+1} \right)} \Gamma_{\rm d} t\right]}.
\end{align}
For $J_{\rm prop}/\Gamma_{\rm d}  \ll 1$, we have that
\begin{equation}
    \hat{H}_{0}
    =
    -i \hat{n}^{\alpha}_L
    ,
    \quad
    \hat{H}^{'}
    =
    -\frac{J_{\rm prop}}{\Gamma_{\rm d}}
    \sum_{\ell=1}^L (\hat{\alpha}_{\ell}^\dagger \hat{\alpha}^{}_{\ell+1}+\textrm{h.c.} ).
\end{equation}
Although the zero order eigenvectors are degenerated, we can express them as the eigenvectors of $\hat{H}^{'}$ in the subspace corresponding to the first $L-1$ sites, so that
\begin{align}
    \ket{\psi_\ell^0}=\frac{1}{\sqrt{\mathscr{N}_{\ell}^{'}}}&\left( \sin{\left( \frac{\ell\pi}{L} \right)}, \sin{\left( \frac{2\ell \pi}{L} \right)},   \cdots,  \sin{\left( \frac{\ell^2 \pi}{L} \right)} \right. \nonumber \\
    &\quad \left.,  \cdots,  \sin{\left( \frac{(L-1)\ell \pi}{L} \right)},   0 \right)^T, \ E_\ell^{(0)}=0, \\
    \ket{\psi_L^0}=(0,  0,  & \cdots,  0,  \cdots,  0,   1 )^T, \quad  E_L^{(0)}=-i,
\end{align}where $\ell=1,2..,L-1$ and $\mathscr{N}_\ell^{'}=\sum_{k=1}^{L-1} \sin^2{\left( \frac{k\ell \pi}{L} \right)}$. First order corrections to the energies are real, see. Eq.~\eqref{L3_diss_firstcorrection} and not relevant for our purposes. For the second order correction, we take into account that
\begin{equation}
    \braket{\psi_i^0|\hat{H}^{'}|\psi_j^0}=0, \quad \forall i\neq j=1,2,...,L-1,
\end{equation}
such that we need to consider only one term, so 
\begin{align}
    E_\ell^{(2)}=- \left( \frac{J_{\rm prop}}{\Gamma_{\rm d} } \right)^2\frac{\sin^2{\left( \frac{\ell (L-1) \pi}{L} \right)}}{\mathscr{N}_\ell^{'}}i.
\end{align}
Then the norm evolves as
\begin{align}
    \mathscr{N}_{\psi,\rm low} \approx& \sum_{\ell=1}^{L-1} \frac{\sin^2{\left( \frac{\ell \pi}{L} \right)} }{\sum_{k=1}^{L-1} \sin^2{\left( \frac{k\ell \pi}{L} \right)}} \nonumber \\
    &\times \exp{\left[-\frac{2 J_{\rm prop}^2}{ \Gamma_{\rm d}} \frac{\sin^2{\left( \frac{(L-1) \ell \pi}{L} \right)} }{\sum_{k=1}^{L-1} \sin^2{\left( \frac{k\ell \pi}{L} \right)}} t\right]}.
\end{align}
We finally obtain the norm in the two limits
\begin{align}
    \mathscr{N}_{\psi,\rm low} \approx 
    \begin{cases}
    \sum_{\ell=1}^L \frac{\sin^2{\left( \frac{\ell \pi}{L+1} \right)} }{\sum_{k=1}^L \sin^2{\left( \frac{k\ell \pi}{L+1} \right)}} &
    \\
    \times \exp{\left[-\frac{4}{L+1}\sin^2{\left( \frac{\ell L \pi}{L+1} \right)} \Gamma_{\rm d} t\right]}, \frac{\Gamma_{\rm d}}{J_{\rm prop}} \ll 1 \\
    \sum_{\ell=1}^{L-1} \frac{\sin^2{\left( \frac{\ell \pi}{L} \right)} }{\sum_{k=1}^{L-1} \sin^2{\left( \frac{k\ell \pi}{L} \right)}}& \\
    \times \exp{\left[-\frac{2 J_{\rm prop}^2}{ \Gamma_{\rm d}} \frac{\sin^2{\left( \frac{(L-1) \ell \pi}{L} \right)} }{\sum_{k=1}^{L-1} \sin^2{\left( \frac{k\ell \pi}{L} \right)}} t\right]}, \frac{\Gamma_{\rm d}}{J_{\rm prop}}\gg 1 
    \label{L_diss_prop_noedgelocalization}
    \end{cases}
\end{align}
Finally, we summarize in the Table~\ref{table_summary_norm} all the results for the norm evolution in the leakage propagation case. Taking into account the general $L$ without borders, we observe a tendency where increasing the size of the array increases the leakage population removal times and reduces the value of the low optimal measurement rate itself. The differences between the cases with and without borders for $L=3$ indicate also that the edge-localization effect has a similar tendency.
\begin{table*}
\caption{Summary of the results for the evolution of the norm in the case of leakage propagation interacting with dissipation at the last site.}
\begin{tabular}{ | m{8em} | m{7cm}| m{7.5cm} | } 
  \hline
   & $\frac{ \Gamma_{\rm d}}{J_{\rm prop}} \ll 1$ & $\frac{ \Gamma_{\rm d}}{J_{\rm prop}} \gg 1$ \\ 
  \hline
  $L=2$ & $e^{-\Gamma_{\rm d} t}$ & $e^{-\frac{2 J_{\rm prop}^2}{\Gamma_{\rm d}} t}$ \\ 
  \hline
  $L=3$ & $\frac{1}{2}e^{-\Gamma_{\rm d} t}+\frac{1}{6}e^{-\frac{1}{3}\Gamma_{\rm d} t}+\frac{1}{3}e^{-\frac{2}{3}\Gamma_{\rm d} t}$ & $0.3e^{-0.7\frac{2 J_{\rm prop}^2}{\Gamma_{\rm d}} t}+0.7e^{-0.3\frac{2 J_{\rm prop}^2}{\Gamma_{\rm d}} t}$ \\ 
  \hline
  $L=3$ (no borders) & $\frac{1}{2}e^{-\Gamma_{\rm d} t}+\frac{1}{2}e^{-\frac{1}{2}\Gamma_{\rm d} t}$ & $e^{-\frac{J_{\rm prop}^2}{\Gamma_{\rm d}} t}$ \\
  \hline
  $L$ (no borders) & $\sum_{\ell=1}^L \frac{\sin^2{\left( \frac{\ell \pi}{L+1} \right)} }{\sum_{k=1}^L \sin^2{\left( \frac{k\ell \pi}{L+1} \right)}} \exp{\left[-\frac{4}{L+1}\sin^2{\left( \frac{\ell L \pi}{L+1} \right)} \Gamma_{\rm d} t\right]}$ & $\sum_{\ell=1}^{L-1} \frac{\sin^2{\left( \frac{\ell \pi}{L} \right)} }{\sum_{k=1}^{L-1} \sin^2{\left( \frac{k\ell \pi}{L} \right)}} \exp{\left[-\frac{2 J_{\rm prop}^2}{ \Gamma_{\rm d}} \frac{\sin^2{\left( \frac{(L-1) \ell \pi}{L} \right)} }{\sum_{k=1}^{L-1} \sin^2{\left( \frac{k\ell \pi}{L} \right)}} t\right]}$ \\
  \hline
\end{tabular}
\label{table_summary_norm}
\end{table*}

\subsection{Leakage disintegration}
Considering the full space expressed in the basis $\ket{20}, \ket{11}, \ket{02}$, the effective Hamiltonian is given by
\begin{equation}
    \hat{H}_{\rm{NJ}}
    =
    \begin{pmatrix}
    -\bar U & \sqrt{2}J & 0\\
    \sqrt{2}J & -i\frac{\Gamma_{\rm d}}{2} & \sqrt{2}J \\
    0 & \sqrt{2}J &  -\bar U-i\Gamma_{\rm d}
    \end{pmatrix}.
    \label{H_eff_splitting_diss}
\end{equation}
By considering $\hat{H}_{\rm{eff}}=U(\hat{H}_0+\hat{H}^{'}) $, we solve this system perturbatively in the limit of $J/\bar U \ll 1$, obtaining the eigenvectors and eigenvalues
{\allowdisplaybreaks
\begin{align}
    \ket{\psi_1^0}=&(1 , 0 , 0 )^T,\quad E_1^{(0)}=-1, \\
    \ket{\psi_2^0}=&(0 , 1 , 0 )^T,\quad E_2^{(0)}=-i\frac{\Gamma_{\rm d}}{2\bar U}, \\
    \ket{\psi_3^0}=&(0 , 0 , 1 )^T,\quad E_3^{(0)}=-1-i\frac{\Gamma_{\rm d}}{\bar U},
\end{align}}where $\ket{\tilde{\psi}_n^0}=\ket{\psi_n^0} $ and $E_n^{(0)}=(\tilde{E}_n^{(0)})^*$ for every~$n$. The first order correction to the energy is $E_n^{(1)}=0$ for every $n$, and the second order corrections are given by
{\allowdisplaybreaks
\begin{align}
    E_1^{(2)}=&-\left(\tfrac{J}{\bar U}\right)^2 \frac{2\left(1+i\frac{\Gamma_{\rm d}}{2\bar U} \right)}{1+ \left( \frac{\Gamma_{\rm d}}{2\bar U}\right)^2}, \\ 
    E_2^{(2)}=&\left(\tfrac{J}{\bar U}\right)^2 \frac{4}{1+ \left( \frac{\Gamma_{\rm d}}{2\bar U}\right)^2}, \\
    E_3^{(2)}=&-\left(\tfrac{J}{\bar U}\right)^2 \frac{2\left(1+i\frac{\Gamma_{\rm d}}{2\bar U} \right)}{1+ \left( \frac{\Gamma_{\rm d}}{2\bar U}\right)^2}.
\end{align}}Substituting into~\eqref{norm_evolution}, we finally have the expression
\begin{align}
    \mathscr{N}_{\psi,\rm high} \approx {\rm e}^{2\textnormal{Im}(E_1)t } &= \exp\left(-\frac{t}{T_\star^{\rm{high}, \ \rm{d}}}\right) \notag \\ &= \exp\left( -\frac{8J^2 \Gamma_{\rm d}}{4\bar U^2+ \Gamma_{\rm d}^2}t\right).    
    \label{norm_split_dissipation}
\end{align}
We get a high optimal rate $\Gamma_{\rm d}^{\rm{high}} \approx 2\bar U$, and a decay time for the leakage population at the high optimal rate of $T_\star^{\rm{high}, \ \rm{d}}(\Gamma_{\rm d}=\Gamma_{\rm d}^{\rm{high}})\approx\bar U/2J^2$. Although, we do not perform any analysis for $L>2$ for the high optimal rate, we expect an increase in the leakage population removal time with an increase in chain length, as the difference with the leakage propagation case lies solely in the last transmon, i.e., disintegration instead of propagation. However, in the numerical simulations for $L=3$, see Fig.~\ref{fig:leakage_works} in the main paper, we observe an increase in the high optimal dissipation rate value with respect to the $L=2$ analytical value found here. We hypothesize that this could be explained because of the edge-localization effect, which reduces the rate of the leakage propagation increasing the rate of leakage disintegration.

\subsection{Qubit subspace dynamics}
In this subsection, we study the effect of dissipation in the qubit subspace, i.e., how dissipation at the last site affect single excitations at the first site. The effective Hamiltonian is given by
\begin{equation}
    \hat{H}_{\rm{NJ}}
    =
    \sum_{\ell=1}^L \left[\omega_{\ell} \hat{n}_\ell+J_{\ell} \left(\hat{a}_{\ell}^\dagger \hat{a}^{}_{\ell+1}+\textrm{h.c.} \right)\right]-i\frac{\Gamma_{\rm d}}{2}\hat{n}_L.
\end{equation}
For the simplest case of a non-disordered $L=2$ transmons, we exactly diagonalize system avoiding the exceptional point $\Gamma_{\rm d}=4J$. For $\Gamma_{\rm d}<4J$, we have
\begin{align}
    \mathscr{N}_\psi=&\frac{4e^{-\frac{\Gamma_{\rm d}}{2} t}}{16-\frac{\Gamma^2_{\rm d}}{J^2}} \left[ 4-\frac{\Gamma^2_{\rm d} }{4J^2} \cos{\left(\frac{t}{2} \sqrt{16J^2-\Gamma_{\rm d}^2}\right)} \right. \nonumber \\
    &\left.+\frac{\Gamma_{\rm d}}{J} \sqrt{1-\frac{\Gamma^2_{\rm d} }{16J^2}}\sin{\left( \frac{t}{2}\sqrt{16J^2-\Gamma_{\rm d}^2}\right)} \right],
\end{align}
and for $\Gamma_{\rm d}>4J$,
\begin{align}
    \mathscr{N}_\psi=&\frac{4e^{-\frac{\Gamma_{\rm d}}{2} t}}{\frac{\Gamma^2_{\rm d}}{J^2}-16} \left[-4+\frac{\Gamma^2_{\rm d} }{4J^2} \cosh{\left(\frac{t}{2} \sqrt{\Gamma_{\rm d}^2-16J^2}\right)} \right. \nonumber \\
    &\left.+\frac{\Gamma_{\rm d}}{J} \sqrt{\frac{\Gamma^2_{\rm d} }{16J^2}-1}\sinh{\left( \frac{t}{2}\sqrt{\Gamma_{\rm d}^2-16J^2}\right)} \right].
\end{align}
To have a simpler expression for interpretation and comparison purposes, we also solve the system perturbatively: for $ 2J/\Gamma_{\rm d}\ll 1$, we obtain $ \mathscr{N}_\psi \approx e^{-\frac{4J^2}{ \Gamma_{\rm d}}t}$, and for $ \Gamma_{\rm d}/2J \ll 1$ we obtain $ \mathscr{N}_\psi \approx e^{-\Gamma_{\rm d} t}$. Therefore, we have the approximate expression $\mathscr{N}_\psi \approx \exp{\left[-\frac{4J^2\Gamma_{\rm d}}{8J^2+\Gamma_{\rm d}^2}t \right]}$, where we get the low optimal rate $\Gamma_{\rm d}^{\rm low} \approx 2\sqrt{2}J$. 

When there is disorder $\omega_1 \neq \omega_2$, we can solve the Hamiltonian perturbatively, for $J/\bar \omega \ll 1$, where $\omega_\ell=\bar \omega+\delta \omega_\ell$. We have that
{\allowdisplaybreaks
\begin{align}
    \hat{H}_0
    =&
    \sum_{\ell=1}^2 \left(1+\frac{\delta \omega_\ell}{\bar \omega}\right) \hat{n}_\ell-i\frac{\Gamma_{\rm d}}{2 \bar  \omega}\hat{n}_2
   ,
    \\
    \hat{H}^{'}
    =&
    \frac{J}{\bar \omega}
    \sum_{\ell=1}^2\left(\hat{a}_{\ell}^\dagger \hat{a}^{}_{\ell+1}+\textrm{h.c.} \right),
\end{align}}
and the eigenvectors and eigenvalues are
{\allowdisplaybreaks
\begin{align}
    \ket{\psi_1^0}=&\begin{pmatrix}1 \\ 0   \end{pmatrix},\quad E_1^{(0)}=1+\frac{\delta \omega_1}{\omega}, \\
    \ket{\psi_2^0}=&\begin{pmatrix}0 \\ 1   \end{pmatrix},\quad E_2^{(0)}=1+\frac{\delta \omega_2}{\omega}-i\frac{\Gamma_{\rm d}}{2\omega},
\end{align}}expressed in the basis $ \ket{10}, \ket{01} $. The first order correction for the energy is $E_n^{(1)}=0$, for every $n$, and the second order correction are
{\allowdisplaybreaks
\begin{align}
    E_1^{(2)}= \frac{J^2}{\bar \omega} \frac{\left[\left( \delta \omega_1-\delta \omega_2 \right)-i\frac{\Gamma_{\rm d}}{2} \right]}{\left( \delta \omega_1-\delta \omega_2 \right)^2+ \left( \frac{\Gamma_{\rm d}}{2}\right)^2}, \\
    E_2^{(2)}=\frac{J^2}{\bar \omega} \frac{\left[\left( \delta \omega_2-\delta \omega_1 \right)+i\frac{\Gamma_{\rm d}}{2} \right]}{\left( \delta \omega_2-\delta \omega_1 \right)^2+ \left( \frac{\Gamma_{\rm d}}{2}\right)^2}.
\end{align}}For the dissipation case, we cannot calculate $T_1^{\rm{d}}$ and $T_2^{\rm{d}}$ in the expressions of $\braket{\hat{n}_1}$ and $\braket{\hat{\rho}_+}$ since $p_{\rm{d}}(t)=1-\mathscr{N}_\psi$ is the probability of a dissipation event to happen. However, we can calculate the typical times $\tau_1$ and $\tau_2$ in the expressions $p_{\rm{d}}(t)=1-e^{-t/\tau_1}$ and $p_{\rm{d}}(t)=1-e^{-t/\tau_2}$, related to $\braket{\hat{n}_1}$ and $\braket{\hat{\rho}_+}$ respectively. For obtaining $\tau_1$, we consider the initial state $\ket{10}$ such that
\begin{equation}
    \mathscr{N}_\psi \approx e^{2\textnormal{Im}(E_1)t}  \approx \textnormal{exp}\left(- \frac{J^2 \Gamma_{\rm d}}{\left(\delta \omega_1 -\delta \omega_2 \right)^2+ \left( \frac{\Gamma_{\rm d}}{2}\right)^2}t\right).
\end{equation}
For obtaining $\tau_2$, we consider the initial state $\frac{1}{\sqrt{2}}\left(\ket{1}+\ket{0}\right)\ket{0}$ such that
\begin{align}
    \mathscr{N}_\psi &\approx \frac{1}{2} \left(e^{2\textnormal{Im}(E_1)t}+1\right) \nonumber \\
    &\approx\frac{1}{2} \left[\textnormal{exp}\left(- \frac{J^2 \Gamma_{\rm d}}{\left(\delta \omega_1 -\delta \omega_2 \right)^2+ \left( \frac{\Gamma_{\rm d}}{2}\right)^2}t\right)+1\right].
\end{align}
Although $\tau_1$ and $\tau_2$ refers to the probability of removing one excitation, we can infer the values of $T_1^{\rm{d}}$ and $T_2^{\rm{d}}$ for $\braket{\hat{n}_1}$ and $\braket{\hat{\rho}_+}$, respectively. For the case of $T_1^{\rm{d}}$, we can assume that removing one excitation is directly related with the decay of the initial state, while for $T_2^{\rm{d}}$, we can gain insight from Eq.~\eqref{T2_fb} and understand that removing one excitation changes the initial state by half, such that
\begin{equation}
    T_1^{\rm{d}}\approx\frac{4\left(\delta \omega_1 -\delta \omega_2 \right)^2+ \Gamma_{\rm d}^2}{4J^2 \Gamma_{\rm d}}, \quad
    T_2^{\rm{d}}\approx\frac{4\left(\delta \omega_1 -\delta \omega_2 \right)^2+ \Gamma_{\rm d}^2}{2J^2 \Gamma_{\rm d}}
    \label{T1_T2_diss}.
\end{equation}
The optimal rate for removing the qubit subspace population is $\Gamma_{\rm d}^{[T_1]}=\Gamma_{\rm d}^{[T_2]}=2|\delta \omega_1 -\delta \omega_2|$. Therefore, for protecting the qubit subspace, we need to choose a value of $\Gamma_{\rm d}$ far from this optimal rate.

\begin{figure}
    \centering
    \includegraphics[width=1\linewidth]{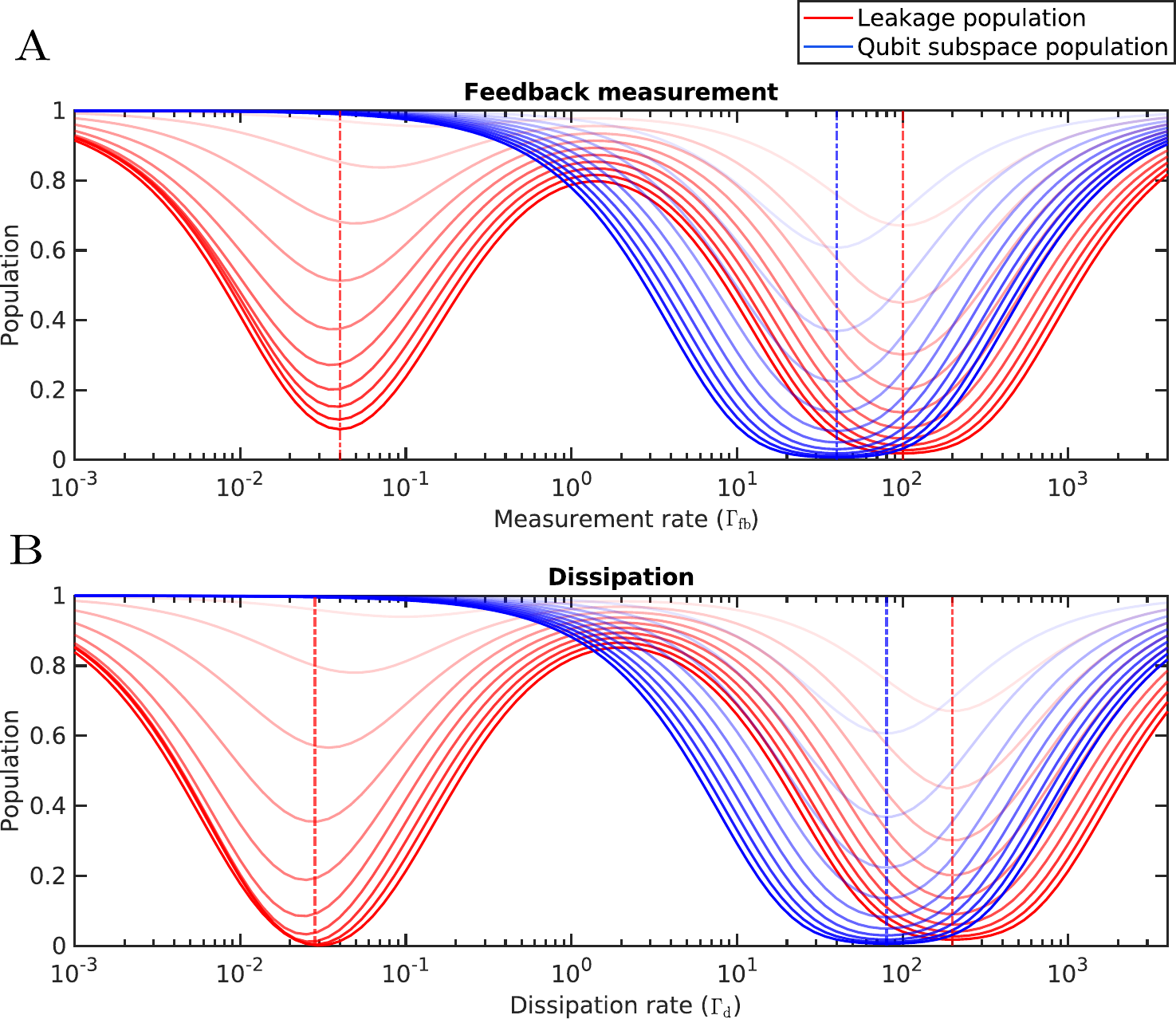}
    \caption{Dynamics of the leakage and qubit subspace populations for different rates for last-reset by feedback measurements (A) or dissipation (B). Vertical dashed lines indicate optimal rates obtained analytically: $\Gamma_{\rm fb}^{\rm{low}}\approx2 J_{\rm prop}$, $\Gamma_{\rm fb}^{\rm{high}} \approx \bar U$, $\Gamma_{\rm d}^{\rm{low}}\approx \sqrt{2} J_{\rm prop}$, and $\Gamma_{\rm d}^{\rm{high}}\approx 2 \bar U$ for leakage population, and $\Gamma_{\rm fb}^{[T_1]} = \Gamma_{\rm fb}^{[T_2]}\approx |\omega_1-\omega_2|$ and  $\Gamma_{\rm d}^{[T_1]}=\Gamma_{\rm d}^{[T_2]} \approx 2|\omega_1-\omega_2|$ for qubit subspace population for feedback measurement and dissipation respectively. The values are taken from numerically solved the master equation using the parameters $U_1/J=60$, $U_2/J=140$, $\omega_1/J=980$, $\omega_1/J=1020$ Each line represents a time instance, starting from $t_0J=0$ (faint) to $t_FJ=200$ (bold) in steps of $\Delta tJ=20$. }
    \label{fig:scheme_result}
\end{figure}

\subsubsection{General $L$}
Interestingly, we can obtain an expression for arbitrary LRU sizes. Proceeding as above, for the case $J/\bar \omega \ll 1$ we have that
{\allowdisplaybreaks
\begin{align}
    \hat{H}_0
    =&
    \sum_{\ell=1}^L \left(1+\frac{\delta \omega_\ell}{\bar \omega}\right) \hat{n}_\ell-i\frac{\Gamma_{\rm d}}{2 \bar  \omega}\hat{n}_L
   ,
    \\
    \hat{H}^{'}
    =&
    \frac{J}{\bar \omega}
    \sum_{\ell=1}^L \left(\hat{a}_{\ell}^\dagger \hat{a}^{}_{\ell+1}+\textrm{h.c.} \right),
\end{align}}Assuming that $\delta \omega_1 \neq \delta \omega_\ell $ for all $\ell=2,...,L$ we can apply non-degenerate perturbation theory for the eigenvectors and eigenvalues
{\allowdisplaybreaks
\begin{align}
    \ket{\psi_\ell^0}=&\ket{\ell},\quad E_\ell^{(0)}=1+\frac{\delta \omega_\ell}{\bar \omega}, \\  
    \ket{\psi_L^0}=&\ket{L},\quad E_L^{(0)}=1+\frac{\delta \omega_{L}}{\bar \omega}-i\frac{\Gamma_{\rm d}}{2\bar \omega},
\end{align}}where $\ell=1,2,\dots,L-1$. We consider only initial states localized at the first site $\ell=1$. Since the only states that contribute to reduce the norm are those with imaginary terms in their energies, see. Eq.~\eqref{norm_evolution}, for an initial excitation localized at $\ell=1$ we just need to calculate the following term corresponding to the $2(L-1)$-th order correction of the energy
\begin{align}
    E_1^{[2(L-1)]}&=(E_1^{(0)}-E_{L}^{(0)})\prod_{i=1}^{L-1} \frac{\braket{\psi_i^0|\hat{H}^{'}|\psi_{i+1}^0}\braket{\psi_{i+1}^0|\hat{H}^{'}|\psi_{i}^0}}{(E_1^{(0)}-E_{i+1}^{(0)})^2} \nonumber \\
    &=  \frac{J^2\left[\left( \delta \omega_1-\delta \omega_2 \right)-i\frac{\Gamma_{\rm d}}{2} \right]}{\left( \delta \omega_1-\delta \omega_L \right)^2+ \frac{\Gamma^2_{\rm d}}{4}} \prod_{n=2}^{L-1} \frac{J^2}{\left( \delta \omega_1-\delta \omega_n \right)^2}. 
\end{align}
Considering the initial state $\ket{100 \dots 0}$, we have that
\begin{equation}
    \mathscr{N}_\psi\approx e^{2\textnormal{Im}(E_1)t} \approx \textnormal{exp}\left(- \frac{FJ^2 \Gamma_{\rm d}}{\left(\delta \omega_1 -\delta \omega_L \right)^2+ \left( \frac{\Gamma_{\rm d}}{2}\right)^2}t\right),
\end{equation}
and the initial state $\frac{1}{\sqrt{2}}\left(\ket{1}+\ket{0}\right)\ket{00 \dots 0}$,
\begin{align}
    \mathscr{N}_\psi &\approx \frac{1}{2} \left(e^{2\textnormal{Im}(E_1)t}+1\right) \nonumber \\
    &\approx\frac{1}{2} \left[\textnormal{exp}\left(- \frac{FJ^2 \Gamma_{\rm d}}{\left(\delta \omega_1 -\delta \omega_L \right)^2+ \left( \frac{\Gamma_{\rm d}}{2}\right)^2}t\right)+1\right],
\end{align}
where $F=\prod_{n=2}^{L-1} [J^2/\left( \delta \omega_1-\delta \omega_n \right)^2]$. Therefore, adding additional transmons reduce the time decay by a constant depending on the product of disorders with the respect to the initial site. Interestingly, the optimal rate  $\Gamma_{\rm d}$ for disturbing the qubit subspace only depends on the disorder between the initial site and the measured one.

\section{Minimal leakage removal unit}\label{app:minimal_leakage_removal}
In this last section, we summarize in Fig.~\ref{fig:scheme_result} the analytical results of the sections~\ref{app:fb_analytics} and~\ref{app:dissipation_analytics} for the minimal LRU consisting of $L=2$ transmons. For this particular case, we consider a specific set of disorder, without averaging, and under ideal conditions, meaning no dissipation, decoherence, nor non-zero temperature  affecting the system. We see that the best option is to choose $\Gamma_{\rm d}^{\rm{low}}$ as the strategy for removing the leakage population. This is because: i) dissipation is a passive method, ii) $\Gamma_{\rm d}^{\rm{low}}$ represents the optimal rate that is farthest from the rates that affect the qubit subspace population $\Gamma_{\rm d}^{[T_1]}$ and $\Gamma_{\rm d}^{[T_2]}$, and iii) $\Gamma_{\rm d}^{\rm{low}}$ achieves the fastest removal of the leakage population.

Note that for this minimal LRU to efficiently removes the leakage population without affecting the qubit subspace significantly, it is crucial that we set the disorder in advance, without relying on any distribution, so that the results discussed here can be applied to cases where $U_1$ and $U_2$ are known. In this particular conditions, the minimal LRU is quite efficient for three reasons: (i) the leakage population reaches the measured/dissipated transmon faster, (ii) the leakage population is less de-localized along the transmons array, and (iii) the leakage population propagation is fully resonant since there is no edge-localization effect. Note that the qubit subspace is always more affected than in $L>2$ cases; by carefully tuning the parameters to minimize the errors $T_1$ and $T_2$, the minimal LRU works efficiently while requiring fewer elements and thus reducing noise sources.

\bibliography{references}

\end{document}